# An Introduction to Biomolecular Simulations and Docking

Cameron Mura[1*] & Charles E. McAnany[1]

**Author affiliations**:

[1] University of Virginia
Department of Chemistry
Charlottesville, VA 22904, USA

*correspondence can be addressed to CM:
+1 434 924 7824 (tel)
+1 434 924 3710 (fax)
cmura@muralab.org

**Abstract**:

The biomolecules in and around a living cell – proteins, nucleic acids, lipids, carbohydrates – continuously sample myriad conformational states that are thermally accessible at physiological temperatures. Simultaneously, a given biomolecule also samples (and is sampled by) a rapidly fluctuating local environment comprised of other biopolymers, small molecules, water, ions, etc. that diffuse to within a few nanometers, leading to inter-molecular contacts that stitch together large supramolecular assemblies. Indeed, all biological systems can be viewed as dynamic networks of molecular interactions. As a complement to experimentation, molecular simulation offers a uniquely powerful approach to analyze biomolecular structure, mechanism, and dynamics; this is possible because the molecular contacts that define a complicated biomolecular system are governed by the same physical principles (forces, energetics) that characterize individual small molecules, and these simpler systems are relatively well-understood. With modern algorithms and computing capabilities, simulations are now an indispensable tool for examining biomolecular assemblies in atomic detail, from the conformational motion in an individual protein to the diffusional dynamics and inter-molecular collisions in the early stages of formation of cellular-scale assemblies such as the ribosome. This text introduces the physicochemical foundations of molecular simulations and docking, largely from the perspective of biomolecular interactions.

## Contents

## 1. **Introduction**

Molecular biology is highly dynamical in nature, contrary to what may be implied by the static illustrations of proteins, nucleic acids, and other biomolecular structures printed in textbooks. Life occurs above absolute zero, and the biomolecular components in and around a cell – proteins, nucleic acids, lipids, carbohydrates – are continuously sampling, via *intra*-molecular interactions, the myriad conformational states that are thermally accessible at physiological temperatures. Simultaneously, a given biomolecule also samples (and is sampled by) a rapidly fluctuating local environment comprised of other biopolymers, small molecules, water, ions, etc. that diffuse to within a few nanometers, leading to *inter*-molecular interactions and the formation of supramolecular assemblies [1-6]. These intra- and inter-molecular contacts are governed by the same physical principles (forces, energetics) that characterize individual molecules and inter-*atomic* interactions, thereby enabling a unified picture of the physical basis of molecular interactions from a small set of fundamental principles [7-12]. From just a few physical laws, and several plausible assumptions, describing covalent and non-covalent (*non-bonded*) interactions and their relative magnitudes, much can be learnt about molecular interactions and dynamics as the means by which proteins fold into thermodynamically stable 'native' structures [13-15], bind other proteins or small molecules to trigger various cellular responses [16], act as allosteric enzymes [17-20], participate in metabolic pathways and regulatory circuits [21, 22], and so on — in short, all of cellular biochemistry.

Computational approaches are well-suited to studies of molecular interactions, from the intra-molecular conformational sampling of individual proteins (such as membrane receptors [23] or ion channels [24]) to the diffusional dynamics and inter-molecular collisions that occur in the early stages of formation of cellular-scale assemblies (such as a neuronal dendritic spine [25, 26]). To study such phenomena, two major lineages of computational approaches have developed in molecular biology: physics–based methods (often referred to as *simulations*) and informatics–based approaches (often termed the *data-mining* or *machine learning* approach to knowledge extraction via statistical inference). An advantage of the former approach is its physical *realism* [11], while an advantage of the latter approach is its potential to illuminate phylogenetic relationships and *evolutionary features* [27, 28]. This primer focuses on the simulation of biopolymers and molecular interactions as physical processes; introductory texts on bioinformatic approaches are available (e.g., Jones & Pevzner [29]).

## 2. **Motivation for computational approaches**

### 2.1. Molecular interactions in context: Biomolecular structure, function, & dynamics

Life is necessarily dynamic, and it is well-established that the three-dimensional structure and dynamics of a biopolymer link its sequence to its function: A specific sequence of amino acids spontaneously folds into a particular 3D shape which, together with the dynamical properties of that structure, give rise to the evolutionarily conserved biochemical functions associated with the protein sequence [5]. However, it is becoming increasingly clear that biomolecular function is also defined *contextually*, in terms of the ligands and other biopolymers with which a biomolecule characteristically interacts (Fig 1A). Consider a biopolymer such as a 150-amino acid, two-domain protein, denoted '$\mathcal{P}$' (e.g., the kinase in Fig 1C). Imagine tracking, with high temporal (≈ ns) and spatial (≈ nm) resolution, a particular copy of $\mathcal{P}$ in a given cell (call it $\mathcal{P}_1$). The physiological activities of $\mathcal{P}_1$ stem from its 3D structure and intrinsic flexibility (*conformational dynamics* within and between its two domains), together with *(i)* the influence of extrinsic factors such as $\mathcal{P}_1$'s chemical environment (redox potential, pH, ionic strength, etc.), and *(ii)* the set of *molecular interactions* in which $\mathcal{P}_1$ engages at any single instant with copies of itself and other biopolymers ($\mathcal{Q}$, $\mathcal{R}$, …), ligands, etc. This set of molecular contacts $\mathcal{P}_1 \cdots \{\mathcal{Q}, \mathcal{R}, \ldots\}$ can rapidly change, even on the timescale of the dozens of nanoseconds that elapse while $\mathcal{P}_1$ diffuses ≈ 20 Å at room temperature. Yet even this simple picture has already incorporated flawed assumptions: It is now appreciated that the cytoplasm of a cell is a viscous medium that is densely crowded with biopolymers and other solutes with which molecular interactions occur (Fig 1A) [30, 3, 31, 32], making it inaccurate to model diffusion in such an environment as that in pure water [33]. Regardless of such current limits on our understanding, this crowded and inherently dynamic environment of the cellular interior is one reason why molecular interactions and dynamics pervade biology: Biopolymers *fold* into native 3D conformations; monomers *self-assemble* into higher-order structural units that often are the functional entities (e.g., an oligomeric enzyme with a composite active site at a subunit interface [34, 35]); ions *traverse* the pores of membrane channels [24, 36, 37]; motor proteins and other factors *diffuse* along one-dimensional tracks (DNA, cytoskeletal filaments, etc. [38, 39]), and so on. All of these dynamical processes involve the formation and dissociation of molecular contacts that vary greatly in type, number, and duration.

### 2.2. Simulation as a complement to experimentation

Experimentation, computation, and theory are highly complementary. Experimental data are *real*, but unambiguous results demand a flawless set of control experiments, and even then the results are generally not readily (*directly*) interpretable at an atomic or molecular level; understanding and knowledge emerge gradually, via efforts to interpret experimental data in terms of an underlying theoretical model (e.g., fitting ligand-binding data to chemical equilibria equations and isotherms [40]). With molecular simulations and other forms of computation, virtually any imaginable approach can be devised, implemented, and then applied in order to gain insight for nearly any biomolecular system, at potentially ultra-high resolution in terms of length– (atomic) and time– (sub-ps) scales. However, the degree of correctness and realism is not always clear due to the assumptions, limited sampling, etc. that make the calculations feasible in the first place (precision is more readily assessed than accuracy, especially with computational results) [41-44]. There is no substitute for experimental data, and computational results may be best viewed as more predictive and interpretative than conclusive; together, computation and experimentation can aid the testing and development of coherent theories for the mechanism of a biomolecular phenomenon.

Simulation approaches are especially well-suited to studies of biomolecular structure and dynamics, for reasons that range from conceptual to practical. Conceptually, many computational methods have developed out of the same physical theories (usually statistical mechanics) used to describe biopolymer structure and thermodynamics [45, 11], making computational approaches the natural bridge between experimental data and the models (theories) used to interpret such data [46-48]. In practical terms, two distinct types of issues arise. The first issue is true for all bio-systems: Some experimental methods are inherently limited for certain types of questions for any biomolecular system. The second issue is true of all experimental methods: Some biomolecular systems are experimentally less tractable than others, with the nature of the experimental limitation depending on the precise question. As an example, consider the problem of extracting information about the dynamics of a protein•ligand complex at both atomic resolution and over the potentially relevant ns ↔ ms timescales (Fig 2). Crystallography is not readily applied to this problem because a protein•ligand crystal structure is a spatially and temporally averaged model, the averages being taken over more than $10^{12}$ unit cells (a conservative estimate, for μm-sized crystals of typical cell dimensions) and timespans greater than hundreds of milliseconds (a conservative estimate, for exposure with high-brilliance synchrotron X-rays). The development of time-resolved diffraction approaches [49, 50] is an active area of research that can benefit from simulation approaches [51] as well as new experimental capabilities [52]. Solution-state NMR relaxation measurements offer another experimental methodology to study dynamics, but this approach can be hampered by fundamental timescale issues and by the need to fit data to *a priori* assumptions about motional modes (see, e.g., [53, 54] for discussions).

A basic problem for the diffraction and spectroscopic approaches is that, as structural biology has advanced, many of the systems of contemporary interest are large, dynamic assemblies that may be only transiently stable (e.g., membrane protein complexes [55], the RNA-processing spliceosome [56]). High-resolution crystallographic or NMR studies of such systems are hindered by precisely those features that may be of greatest biological interest — conformational heterogeneity in the population on the timescale of the experiment, dynamical interconversions between stable sub-states (some sub-states may be more ‘druggable’ [57, 58]), and so on. Diffraction studies require well-ordered crystals, and crystallization requires a supersaturated population of molecules or complexes [59]; excessive structural variability among the entities will impede their packing into a geometrically ordered lattice (or, even if they do, the lattice may diffract only poorly due to severe mosaicity or other defects [60]). Similarly, in NMR structure determination [61] the dynamical regions are generally the least well-resolved, and approaches to extract dynamics are beset by potential limitations; for instance, the model-free approach to infer dynamics from spin relaxation measurements assumes decoupling of global (e.g., protein tumbling) and internal (e.g., domain hinge-bending) modes, which is problematic for large-scale, high-amplitude fluctuations such as between two protein domains [54]. Also, electron paramagnetic resonance (EPR) spectral line-shapes can be analyzed to infer ns–scale protein backbone dynamics [62], but this approach alone is not without caveats. Computational approaches such as MD simulation offer an appealing route to exploring molecular flexibility and interactions in full atomic detail, particularly when the desired information is experimentally inaccessible because of these methodological limitations.

### 2.3. Scope of this text

Biomolecular simulation is a vast subject. The remainder of this primer focuses on MD simulations and *in silico* docking, as these are two common computational approaches in the modern biosciences. Also, MD simulations have taken on renewed significance as ultra-long (μs–ms-scale) atomistic simulations are becoming tractable because of advances in hardware, software, and algorithms [63-67, 48, 68, 69]. Intriguing conformational transitions on biologically relevant timescales (μs and beyond; Fig 2) are becoming increasingly accessible using classical MD simulations because of these developments, in addition to a host of 'enhanced sampling' methods that have been under continual development [70, 71, 18, 72]. In what follows, basic concepts (§3) are emphasized rather than practical recipes, with the focus being on MD simulations (§4) and docking (§5). The fundamental principles — conformational sampling, dynamics integrators, force-fields, etc. — appear at the core of most modern computational approaches, including MD and docking. In addition, many interrelated families of techniques derive from the basic MD and docking methodologies, such as coarse-graining [73, 74], simulated annealing structure refinement [75], structure prediction [14, 76, 15], flexible ligand docking [77], protein–protein docking [78-80], and so on.

## 3. **Physical principles, Computational concepts**

The conceptual foundation and practical basis of MD simulations and related approaches, such as Monte Carlo sampling [81, 82], can be appreciated by considering a few key principles. The idea of *energy surfaces* is a unifying physical principle, and *conformational sampling* of the energy landscape is often the computational goal, in both MD and docking. These and related statistical mechanical concepts are described in this section.

### 3.1. Statistical mechanics in a nutshell

#### 3.1.1. Why is it necessary?

Statistical mechanics is the theoretical framework linking the microscopic (atomic-level) properties of a molecule to its thermodynamic properties on bulk/macroscopic scales of, e.g., $10^{23}$ molecules in a vial. To see the need for such a theory, consider an idealized system comprised of a single molecule in complete isolation at $T = 0$ K. Its bond angles, zero-point energy, dipole moment, and other microscopic properties could be computed with reasonable accuracy via quantum mechanics, were the molecule small enough (tens to hundreds of atoms) for the calculations to be feasible at the desired level of QM theory [83]. A molecule under such isolated conditions possesses only a (QM-computable) potential energy, also known as its *internal energy*, $\mathcal{U}$; much of this energy may exist, for instance, by virtue of ring-strain or steric constraints that prevent the molecule from adopting an even lower-energy conformation. Such a system is computationally tractable, but of limited biochemical relevance. Of greater relevance might be the same molecule at finite temperatures (e.g., $T = 310$ K for humans) and on much larger scales, say with $10^{23}$ copies of the compound floating about *in vitro* during a biochemical assay. It is far less straightforward to imagine computing the physical properties (energies, compressibility, etc.) of this bulk system: In addition to the sheer number ($N \approx N_{\mathrm{A}}$) of particles, there is a combinatorial explosion in the number of possible system configurations that must be considered (already $\sim 2^{N_{\mathrm{A}}}$ if there are only two possible states per particle!); there is now kinetic energy to also take into account, which is the thermal energy of a particle by virtue of it being above absolute zero (often denoted as $\mathcal{K}$); there is a continuously dynamical exchange between potential and kinetic energies, the sum of which is the system Hamiltonian ($\mathcal{H} = \mathcal{U} + \mathcal{K}$ for a closed system); there is now a virtual infinitude of potential configurations of system components relative to one another ($\approx N_{\mathrm{A}}^2$ pairwise interactions, to say nothing of three-body and higher-order interactions); there is coupling between the dynamical interactions between particles (inter-molecular dynamics) and the conformational degrees of freedom within individual flexible particles (intra-molecular dynamics); and so on.

#### 3.1.2. Why, and how, does it work?

Despite the complex picture described above, the situation is not hopeless if we take a *statistical* rather than deterministic approach, using probabilistic formulations such as the Boltzmann distribution (Box 2 and below) to describe populations of particles in terms of distributions of microstates and properties (position, velocity, etc.). We compute averages of properties from the statistical distributions, limiting ourselves to bulk scales beyond $N > 10^3$ particles; population sizes less than this are too small. The central pillar of statistical mechanics is a purely numerical property of random variables: *(i)* larger populations have smaller variances in their means (the *law of large numbers*), and *(ii)* large populations of independent random variables tend towards the normal distribution (the *central limit theorem* [84]), with the standard deviation of the mean for a sample ($\sigma_s$) drawn from a population of size $N$ scaling as $\sigma_p/\sqrt{N}$, where $\sigma_p$ is the population standard deviation. As population sizes ap-proach the $N_A$ molecules in a test-tube (e.g., in a calorimetry experiment), the probability density functions (p.d.f.) for any observable/bulk quantity become so strongly spiked that the mean statistical values can be taken as single, well-defined thermodynamic quantities (*entropy*, *free energy*, *etc.*), rather than distributions of values [85, 86]. This asymptotic behavior, $\sigma_s \sim 0$ as $N \to \infty$, is known as the *thermodynamic limit*. Thus, while individual particles in a system of, say, $10^5$ particles may have drastically different individual energies, the mean energy of the system will be essentially a single, well-defined value known as the internal energy, $\langle E \rangle = \mathcal{U}$; the same is true for all bulk properties, such as the heat capacity, entropy, etc.

To illustrate a bulk thermodynamic property in terms of the underlying statistical distributions, consider the entropy ($S$) of a system of $N$ hard spheres. The entropy is a function of the $6N$ particle positions and momenta for the system in discrete microstates $i = 1, 2, 3, \ldots$, and is expressible as a sum over these microstates:

$$S = -k_B \sum_i p_i \ln p_i \qquad (1)$$

In the above, known as the Gibbs entropy formula, $k_B$ is the Boltzmann constant and $p_i$ is the probability of occupation of microstate $i$. An instructive exercise is to consider the Eq 1 summation for the extreme cases of *(i)* a perfectly uniform distribution ($p_1 = p_2 = \cdots = p_N = \frac{1}{N}$) and *(ii)* a singly-spiked distribution ($p_i = 1$ for one $i$); note that, because an infinitude of microstates exist as a continuum in classical dynamics (Fig 3), discrete sums are replaced by integrals in classical statistical mechanics. While the entropy is a measure of the p.d.f. of microstates and is a property of the *ensemble* (Box 2) of particles, it is also a statistical quantity itself – that is, there exists a distribution of entropy values for a system of $N$ particles, too, and that sampling distribution has means ($\langle S \rangle$), variances ($\sigma^2(S)$), and so on. Consider how this distribution of entropy values varies with system size, $N$. The entropy for simple model systems can be computed and plotted for ensembles of size $N = 1, 2, \ldots, N_A$ particles. Extrapolating to $N \approx N_A$ particles, the probability distribution of the entropy p.d.f. becomes infinitely narrow. To see this, consider the exponential growth of the central ($k \approx n/2$) binomial coefficients $\binom{n}{k} = n!/k!\,(n-k)!$ for $n = 1, 2, \ldots, N_A$ coin flip trials, and consider the essentially zero deviation from a 1:1 heads:tails ratio for this series of flips when $n \approx N_A$. In the same way, the distributions of entropy values become so sharply spiked that there are only infinitesimal deviations from the means, $\langle S \rangle$, with $\delta S \approx 0$ as $N \to N_A$. These statistical quantities are precisely the usual thermodynamic properties with which one is familiar, and which can be determined via experimental measurements. Boltzmann showed that for the microcanonical ensemble (Box 2), which corresponds to a constant number of particles ($N$), volume ($V$), and energy ($E$), the system entropy is $S = k_B \ln(\Omega(N, V, E))$, where $\Omega$ is the number of accessible microstates (Fig 3A) and is a function of only $N, V, E$. In this way — as statistical quantities and asymptotic distributions — all the usual thermodynamic potentials take on statistically well-defined (i.e. meaningful) average values on bulk scales.

A biomolecular example to illustrate this general approach would be to consider $10^5$ random conformations of a protein of interest. We can calculate the energy of each of those conformations using the methods of molecular mechanics [8]. Then, given these energies and the Boltzmann distribution, we can evaluate the distribution of conformational states of the protein and also determine the bulk thermodynamic properties of the ensemble [45]. In Maxwell-Boltzmann statistics, the probability of occurrence of state $i$ with associated energy $\varepsilon_i$ ($\langle \varepsilon \rangle = \mathcal{U}$) is:

$$p_i = \frac{e^{-\varepsilon_i/k_B T}}{Z} \qquad (2)$$

where $k_B$ is the Boltzmann constant, $T$ is the absolute temperature, and $Z$ is a normalization constant to ensure that the probabilities sum to unity. This normalization factor is known as the *partition function* [86], and it equals the sum over all microstates $i$. That is, $Z = \sum_i e^{-\varepsilon_i/k_B T}$ in a quantum mechanical formulation; in the classical limit of infinitesimally spaced energy levels, integrals replace discrete sums. Also, this brief introduction does not distinguish between the foregoing molecular partition function and the canonical partition function for an ensemble of $N$ particles at fixed volume and temperature (the $NVT$ ensemble [Box 2]); for the canonical ensemble, the total energy of the system in state $i$, $E_i$, appears in the argument of the exponential. Though the partition function arises as the simple requirement of a valid probability distribution, it is the central link between microscopic properties and macroscopic observables. Indeed, the partition function is a thermodynamic quantity, as seen in the close relationship between the Helmholtz free energy ($A$) and the canonical partition function:

$$A = -k_B T \ln Z \qquad (3)$$

The practical utility of this is the following: Potential energies often can be readily calculated for a given system and, because $Z$ depends on the distribution of potential energies, the results from a computer simulation can provide a description of the partition function. Then, using relations such as Eq 3, the free energy of a system can be calculated directly; recall that free energy $A$ $(= \mathcal{U} - TS)$ gives the maximum amount of work that a closed thermodynamic system can do (in the canonical ensemble, $A$ is minimized at equilibrium). Finally, using the standard Maxwell relations [87], such as the fact that $(\partial T/\partial V)_S = -(\partial P/\partial S)_V$, every thermodynamic quantity (pressure, heat capacity, virial coefficients, etc.) can be derived from this point [88]. Further information on statistical mechanics can be found in Widom's cogent introduction [86] and McQuarrie's [85] comprehensive treatment.

### 3.1.3. What must we consider?

The aforementioned statistical quantities are functions of the probability densities of microstates and their associated energetics. Molecular energetics, in turn, vary with molecular structure (loosely, potential energy) and dynamics (loosely, kinetic energy). The many aspects of molecular structure and dynamics can be synthesized into a coherent framework by considering four basic principles: *(i)* interactions are shaped by the *structural* and *physicochemical properties* of inter-atomic contacts; *(ii)* interactions are *dynamical*, and the macroscopic properties of a system at equilibrium (e.g., ligand-binding free energies [89, 90]) could be exactly computed given full knowledge about the microscopic dynamics of the system's *phase space* (all possible microscopic states, populations of the microstates, transitions between them, etc.; Fig 3); *(iii)* the relative population of different regions of phase space (Fig 3B) define *energy surfaces* for the system of molecular interactions (free energy surface, potential energy surface); and *(iv)* these energy surfaces are *sampled* (basins are populated, barriers are crossed) as the biomolecular system dynamically evolves along a trajectory in phase space. These four considerations — physicochemical interactions, dynamics in phase space, energy surfaces, and conformational sampling — provide a foundation for understanding biomolecular simulations, as described in the remainder of this section.

## 3.2. Physicochemical nature of molecular interactions

Structure and dynamics govern the molecular recognition processes that define the function of a biomolecule. These recognition events involve the non-covalent interactions that occur between the standard chemical functionalities in biopolymers and organic compounds – amines, hydroxyls, carboxylates, amides, aromatic rings, thiols, etc. Viewed hierarchically, a molecular machine as complex as the ribosome (Fig 1B) is simply a specific geometric arrangement of inter-atomic contacts between such functional groups (*structure*), and its *stability* is modulated by the dynamics of these intra- and inter-molecular contacts. Accurate molecular simulations of intra- and inter-molecular contacts require accurate treatment of two basic types of non-bonded interactions: electrostatic interactions [91] and van der Waals (vdW) forces [92].

Electrostatic and vdW interactions differ in their relative magnitudes and in how that magnitude varies with distance between the interacting atoms ($r_{1,2}$). While electrostatic interactions can often be an order of magnitude stronger than vdW energies, both forms of interaction vary greatly with intrinsic factors, such as the types of atoms and bonding involved (element type, hybridization, etc.), as well as extrinsic factors such as the dielectric of the local environment ('$\epsilon$' in the denominator of the Coulombic term in Eq 5 below, which attenuates electrostatic forces). Electrostatic interactions occur between chemical groups that bear formal positive or negative charges (ion pairs, 'salt bridges'), or that contain highly electronegative atomic centers with substantial partial charges (the D–$H^{\delta+}\cdots{}^{-\delta}$A of a hydrogen bond donor/acceptor pair). As a general term for all other (non-electrostatic) forces, van der Waals interactions include forces between two permanent dipoles, a dipole and an induced dipole, or two induced dipoles (the latter are also known as London dispersion forces [92]). Electrostatic forces decay slowly with the distance between interacting centers (Coulombic forces $F \sim 1/r^2$, energies $U_{\mathrm{el}} \sim 1/r$) and are therefore referred to as *long-range*, while vdW forces are considerably more short-ranged. VdW interactions are generally modeled by a Lennard-Jones potential (Eq 5), which contains a $1/r^6$ attractive component that is rooted in the quantum mechanics of London dispersion forces and a $1/r^{12}$ term to capture hard-sphere/exchange repulsion (not physically-based, a numerically convenient expression that can be computed as the square of the $r^{-6}$ term [88]).

Though additional types of interatomic 'forces' are occasionally invoked, such as hydrogen bonding, these are not distinct physicochemical forces. For example, H-bonds, though directional like covalent bonds, are fundamentally electrostatic in nature. Similarly, the hydrophobic effect, which is an important consideration in ligand-binding and drug design, is not a distinct physical force but rather a physical effect that stems from the above forces (electrostatics, vdW) as applied to the properties of liquid $H_2O$ (dipole moment, H-bonding, clathrate-like structures of H-bond networks [93]) under the laws of thermodynamics; the effect is interfacial and is entropically driven (see, e.g., Ch. 8 in [92] and refs. [94-96]). Because electrostatics and vdW interactions are the only fundamental types of intermolecular forces of relevance to biopolymers, molecular mechanics (MM)–based force-field equations are simple in overall functional form (§4.3 and Eq 5, below). These functions, or *potentials* (a term synonymous with force-field), consist of a limited number of bonded and non-bonded terms, usually with all interactions taken as pairwise. The bonded terms represent displacements of bond lengths (stretching), angles (bending), and rotations about covalent bonds (torsional angle); these deviations are modeled as harmonic springs (bonds, angles) or periodic rotation (torsional barrier). The nonbonded terms, which capture all the electrostatic and vdW interactions, correspond to contacts that may be intra-molecular, if the two atoms are in the same molecule (such as between the two domains of the kinase in Fig 1C), or inter-molecular, if the contact occurs between entities in different molecules (such as the antibiotic and the ribosome in Fig 1B). Inclusion of electronic polarizability in FFs is an active area of research, as mentioned below and in Box 4.

In summary, electrostatics and vdW forces are what dictate the structure and energetics of biopolymer folding, assembly, and dynamics, as well as the binding of small molecules, such as antibiotics or other drug compounds, to molecular complexes. Note that even those physiological processes which may not seem non-covalent in character — e.g., electronic transitions accompanying bond formation/rupture, photochemical processes — are still modulated by non-bonded interactions *in vivo*. For instance, the signal transduction cascades underlying vision rely on the covalent attachment of a small polyene known as *retinal* to the protein *opsin*, giving a photo-activatable membrane receptor known as *rhodopsin* [97, 98]. For this to ever occur, the retinal molecule must diffuse to its binding-site in opsin, where it undergoes photon-triggered *cis* → *trans* isomerization in the sterically crowded protein interior; thus, intricate dynamics are at play in each stage of this process. In this sense, molecular dynamics govern virtually all physiological processes, even electronic or photochemical ones.

### 3.3. Dynamical processes and phase space

The formation and stability of molecular interactions are modulated by dynamical processes spanning several decades, ranging from ps-scale rotations of solvent-exposed side-chains near a ligand-binding site to much longer time (≈ ms–μs) *collective motions* that enable allosteric communication (Fig 2; see [99, 100, 58, 101] for examples). For macromolecules, there are three aspects of any dynamical process to consider: *(i)* the *timescale* of the elementary process; *(ii)* the *spatial extent* over which the event occurs; and *(iii)* the *amplitude* of motion. The notion of characteristic times is perhaps the most intuitive of these features: as suggested in Fig 2, various types of dynamical processes occur on timespans that may be narrow and well-defined (e.g., bond vibration), or possibly much broader windows (the collective motions involved in allostery, gated ligand-binding, and biopolymer folding can span several log units [7]). The spatial extent may be small and highly localized (bond vibration, side-chain rotation), or the dynamical process may occur on the length-scale of an entire protein domain, such as in a hinge motion. A similar but not identical concept is the amplitude of oscillatory motion: fluctuations may occur on small or large spatial extents (e.g. two domains of a protein). And, independent of this length-scale, the amplitude itself may correspond to small-scale (≈ Å), high-frequency motion (≈ ns times, corresponding to ≈ GHz in the frequency domain) or larger amplitude (>10 Å), low-frequency (≈ μs-scale) motion. As implied in the foregoing, the frequency and amplitude of motion are often inversely related; this is because a motional mode can be estimated as a normal mode oscillation under a quadratic potential (i.e., harmonic oscillation), for which the mean-square fluctuation for a given amount of energy is $k_{\mathrm{B}}T/\omega_i^2$, where $\omega_i$ is the frequency of mode $i$ [102]. Slow, high-amplitude motions correspond to 'soft' modes that often involve rearrangements of large structural units (helices, sheets, entire domains) and occur over large spatial extents (domains, not side-chains). These long-time dynamics consist of rigid-body motions such as the shearing or twisting of secondary structural elements, the rocking of one domain with respect to another about a hinge, and so on. Low-frequency, high-amplitude motions can be thought of as being 'slower' because they entail extensive sampling of conformational space, wherein the motions of neighboring regions are correlated partly by chance (thermal motions are random), partly by virtue of the pattern of hydrogen-bond connectivity in, say, an α-helix versus a β-strand, and partly by the spatial pattern of other nonbonded interactions between secondary structural elements. These correlated types of motions play key roles in cooperativity and allosteric communication between distant sites in a protein, and also in the fluctuations that modulate the binding of ligands to an effector site [103]. Because long-time dynamics are relatively slow, their time regimes can also overlap the diffusional association of two molecules, which is the first step in molecular recognition.

The preceding discussion implicitly focused on the dynamics of a single biopolymer in isolation. How do the dynamics of a single protein relate to the behavior of a bulk quantity ($N_{\mathrm{A}}$ molecules), as measured in a biochemi-cal assay of, say, ligand–binding affinities? By linking microscopic, atomic-scale dynamics to the macroscopic/thermodynamic properties of a system of molecules, the three concepts of *phase space*, *ensembles*, and *ergodicity* answer this question and provide a complete framework to elucidate how experimental (bulk) quantities relate to the physical and dynamical properties of the system's constituents. Box 2 and the legend to Fig 3 summarize these statistical mechanical concepts. The principle of ergodicity is that an ensemble (bulk) average of some property of a dynamical system asymptotically converges to the time-average of that property, as described below (§3.5). This is the fundamental theoretical justification that allows one to perform MD simulations of single molecules or complexes, versus the computationally unfeasible task of trying to simulate all $\approx N_{\mathrm{A}}$ molecules in a test tube.

### 3.4. A unifying physical picture: Degrees of freedom, energy surfaces

Molecular interactions, A···B, between biopolymers and ligands involve an extraordinary number of degrees of freedom (DoF). A DoF is simply a well-defined parameter that quantifies some property (typically geometric) of a system, where the parameter is free to vary across a range of values independently of other degrees of freedom. Together, all the DoFs define the precise state of a system. For example, a one-dimensional spring at rest is characterized by a specific mechanical equilibrium length, $x_{\mathrm{eq}}$; as the spring executes dynamics in accord with Hooke's law, the length at time $t$, $x(t)$, deviates either as a compression or extension. This deviation $(x - x_{eq})$ is a *translational* DoF of the spring. Analogously, rotation about the central bond in ethane ($\varphi$) is an *angular* DoF, with well-defined bounds of $\varphi \in [0, 2\pi]$. For both the macroscopic spring and microscopic ethane molecule, the energy $E$ (and its negative gradient, the force $\vec{F} = -\nabla E$) is typically some particular function of the DoF: The spring's energy varies quadratically with its sole DoF (Fig 2), defining a parabolic energy surface, while the ethane molecule's potential energy varies periodically with the dihedral $\varphi$ (see the sinusoidal torsion angle term in the FF equation of §4.3). For a system with $n$ DoF, the *energy surface* is simply an $n$-dimensional surface, in $n$+1–dimensional space, giving the energy as a function of the $n$ degrees of freedom [104, 8].

The energy surface concept is entirely generic: Surfaces may correspond to only potential energy terms, as in molecular mechanics, or they may also include thermal energy, thereby corresponding to free energies (as in molecular dynamics). The hyper-dimensional energy surface may be fairly smooth — imagine a simple molecule such as butane *in vacuo* (few DoF). Or, the surface may be corrugated [105, 106], with peaks and valleys of vastly differing magnitude and shape — imagine a protein surrounded by solvent (solvent DoF also would need to be accounted for in computing the energetics of the system). A molecule of $N$ atoms in three-dimensional space has $3N$ degrees of freedom, of which $3N$–6 are vibrational ($3N$–5 if the molecule is linear), and the system's conformational energy surface can be naturally expressed in terms of these $3N$–6 DoF as a vibrational basis set.

Because of its generality, the energy surface offers an integrated physical picture for all aspects of molecular structure, dynamics, thermodynamics, and kinetics. How is this possible? Consider a protein $\mathcal{P}$ and two of its possible states, $\mathcal{P}_{\mathrm{A}}$ and $\mathcal{P}_{\mathrm{B}}$ (e.g., active and inactive states of the protein kinase in Fig 1C). $\mathcal{P}_{\mathrm{A}}$ is a specific 3D structure (*conformation*) that maps to a particular point on the energy surface, and transitions between structural conformers ($\mathcal{P}_{\mathrm{A}}\rightarrow\mathcal{P}_{\mathrm{B}}$, $\mathcal{P}_{\mathrm{A}}\leftarrow\mathcal{P}_{\mathrm{B}}$) occur via dynamical paths (trajectories) along this energy surface. Such transitions persistently occur at finite temperature, assuming any energy barriers between A and B to be surmountable; and, at thermodynamic equilibrium there will be no net change in the relative populations of different regions of the energy surface (valleys, peaks, plateaus). These relative populations reflect macroscopic/thermodynamic energy differences (recall the Boltzmann distribution), while the microscopic details of the transition paths – barrier shapes and heights – dictate the kinetic properties for elementary, single-step transitions in this 'two-state' behavior. The discrete states A and B, corresponding to two basins in the energy surface, can be discrete structural or functional states of protein $\mathcal{P}$ or any dynamical process (A/B may be bound/unbound, folded/unfolded, etc.). Peaks (local maxima) along a pathway from $\mathcal{P}_{\mathrm{A}} \leftrightarrow \mathcal{P}_{\mathrm{B}}$ are transition states, while states A and B themselves are preferentially populated and are referred to as local minima. The depth of a particular basin in a 'funneled' landscape is its enthalpy, while the width of the energy surface near this local minimum reflects the entropy of that state. (Recall that entropy is a measure of the number of thermally accessible states, so a wide/shallow basin corresponds to greater entropy than does a narrow/deep one; this 'entropy/enthalpy compensation' is why the deepest basin is not necessarily the unique global free energy minimum [47].) If the energy surface under consideration is the Gibbs free energy, the relative populations of $\mathcal{P}_{\mathrm{A}}$ and $\mathcal{P}_{\mathrm{B}}$ can be used to compute standard-state free energy differences ($\Delta G°$) for folding, ligand-binding [89, 107], or any other A $\rightleftharpoons$ B process of interest. Again, statistical mechanics is the link between the microscopic dynamics of a single particle on the energy surface and the bulk behavior of an ensemble of particles.

If biomolecular energy surfaces could be fully mapped, we could compute any property of interest for a particular system and its dynamics. However, the sheer number of DoFs for even simple biopolymers leads to an exponentially vast conformational space, making exhaustive exploration of macromolecular energy surfaces an impossible task. The high dimensionality of energy surfaces poses many difficulties, so conformational sampling of the surfaces becomes the crucial computational challenge.

### 3.5. A key computational goal: Conformational sampling

Because we are often concerned with the bulk properties of a system, as determined via experiments, our essential computational goal is to sample molecular conformations across the energy landscape, in accord with a well-defined statistical mechanical ensemble. Thermodynamic equilibrium is generally assumed in such sampling, though this is not strictly necessary; for instance, there exist 'steered' and 'biased' simulation methods that are the computational analog of non-equilibrium, single-molecule 'pulling' experiments [108]. If conformational sampling is done properly – with properly-weighted microstates and sufficient sampling – then we can compute accurate means, deviations, and other statistical values for many types of potentially interesting properties, including *(a)* structural features, such as the radius of gyration; *(b)* thermodynamic quantities, such as entropies and free energies; and *(c)* dynamical properties that supply kinetic/mechanistic insights, such as the correlation time for the motion of a specific loop that 'gates' the binding of ligands to an effector site on a protein [109].

#### 3.5.1. Three types of methods, based on structure, thermodynamics, & kinetics

Conformational sampling approaches can be distinguished from one another based on whether they supply information about problem types *(a)*, *(b)*, or *(c)*, listed above. For instance, some sampling methods focus solely on generating conformations and evaluating their energies, perhaps as trial conformers for structure prediction or for NMR structure determination. In such approaches, which address problem domain *(a)*, the 'energies' can be viewed very generally — not necessarily as physical quantities, but rather as the values of objective functions that quantify the discrepancy between a candidate structure and the experimental data. With sufficiently extensive sampling, optimal agreement between a structural ensemble and experimental data can be achieved by minimizing/maximizing such target functions [75]. The sampling techniques in this class of methodologies – e.g., distance geometry methods, genetic algorithms – are largely heuristic and often are not physically-based, though they can be highly effective ways to sample conformational space from a purely structural perspective (problem type *(a)*). To illustrate the flexibility of these ideas, note that non-physical sampling methods, such as genetic algorithms [8], can be combined with physics-based descriptions of molecular interactions, such as an MM-based force-field, as done in the AUTODOCK software for protein/ligand docking [110].

Turning to the two other classes of approaches, *(b)* and *(c)*, an advantage of physics-based sampling techniques is that they can be used to compute thermodynamic quantities (problem type *(b)*). The two families of such methods are distinguished by whether or not the method aims to *simulate* the underlying dynamics of the system. The first family of approaches, exemplified by Monte Carlo sampling [82], provides correct, Boltzmann-weighted sampling of an ensemble, but does not attempt to simulate the actual microscopic dynamics of the ensemble. Such methods can be used to address questions of structure (*a*, above) and thermodynamics (*b*, above), but not kinetics (*c*, above) [111, 8, 81, 10]. The second family of physics-inspired methods seeks to model – with physical realism – the underlying dynamical processes. These simulation-based approaches, of which MD simulations are a prime example, can supply detailed information on structural (*a*), thermodynamic (*b*), and kinetic (*c*) properties. Simulation methodologies range from a high level of detail, such as all-atom MD incorporating explicit solvent molecules [46], to 'implicit solvent' models [112] that enable more extensive sampling (longer simulations) by treating the solvent as a dielectric continuum (thus reducing the number of DoFs), to further simplified 'coarse-grain' models (e.g., each amino acid modeled as a bead that interacts with other residues under an effective pair potential that has been calibrated for such simulations [73, 113, 114]). MD-based simulation methods can be applied to study the conformational dynamics of single proteins, and even molecular assemblies as complex as the ribosome [115, 116] or as large as an entire HIV capsid [117]. To simulate molecular contacts and diffusional association on long timescales and large spatial domains, Brownian dynamics (BD) methods can be applied [118, 119]. The BD approach typically treats the interacting molecules as rigid; thus, although diffusion-controlled reactions and the long-time behavior of large systems can be simulated, atomically-detailed dynamics are not modelled (see below). To clarify the relationships between various sampling methods, two preva-lent approaches (MD and MC) are compared in Fig 3; further information on conformational sampling and related simulation issues can be found in van Gunsteren *et al.* [47].

#### 3.5.2. Langevin dynamics as a general framework

The MD and BD simulation approaches can both be understood as limiting cases of a single formulation of classical dynamics, namely Langevin dynamics (LD). As described below (§4), the central equation in MD is Newton's second law, $\vec{F} = m\vec{a}$, which describes the classical mechanics of macroscopic systems. The Langevin equation [120, 121, 10] is a phenomenological extension of this law which renders it more generally suitable for dynamic simulations, such as in implicit solvent (e.g., no explicit $H_2O$ molecules, but need to model a stochastic heat bath). In LD, two terms are added to Newton's equation: *(i)* a frictional term that captures dissipative effects, such as frictional drag of solvent molecules on the solute, and *(ii)* a noise term that corresponds to Gaussian-distributed white noise, in order to model the random collisions and 'kicks' between solvent and solute molecules. These terms, which make the Langevin equation a stochastic partial differential equation, are meant to account for the neglected degrees of freedom (e.g., from all the $H_2O$ molecules). The two terms are linked via the fluctuation-dissipation theorem of statistical physics [122, 85] and, because they are both thermal (statistical) in nature, they offer a route to controlling the temperature of a simulation system by adjusting the frictional and collisional coefficients. This is useful because, for instance, many biological simulations are performed in constant-temperature ensembles such as $NPT$ or $NVT$ [123, 124]. The limit of zero frictional coefficient corresponds to purely 'inertial' (Newtonian) dynamics, wherein solvent effects are neglected and the Langevin equation reduces to Newton's second law. Reciprocally, in the 'diffusive' limit of large frictional coefficients, the LD formulation corresponds to more 'random' motion and yields Brownian dynamics.

#### 3.5.3. Sampling and ergodicity

Sampling tasks are exacerbated by two features of energy surfaces: *(i)* their vast dimensionality and *(ii)* their finely nuanced topography, featuring many peaks, valleys, and ridges of greatly varying magnitudes. These two problems are interrelated. Problem *(i)* means that the degree of computational sampling will be quite limited, making it all the more important to sample the most relevant regions of this space; here, 'relevant' is in the sense of low-energy regions, which contribute proportionately more to the equilibrium ensemble average as per their Boltzmann weights. The sampling limitation has motivated the development of 'importance sampling', 'enhanced sampling' approaches, and a host of related algorithms (reviewed in [71]). Obstacle *(ii)* means that, in practice, a simulation may get 'stuck' in a low-lying region of the energy surface, with insufficient thermal inertia to surmount local energy barriers. In such cases, novel or biologically relevant conformational transitions may be completely missed, or sampled an insufficient number of times to enable statistically significant calculation of dynamical properties (lifetimes, mean first passage times, etc.). A general principle for sampling a physical quantity, $Q$, which fluctuates with characteristic time $\tau_Q$, is that the dynamics should be sampled for at least a decade longer than the correlation time [104]; i.e., the simulation length should exceed $10\tau_Q$ if statistically reliable averages are desired. For these reasons, extensive sampling is crucial in MD simulations of biomolecular systems, where interesting transitions often occur on timescales that are quite slow relative to simpler molecular systems.

Getting stuck in a region of conformational space also violates a fundamental axiom of statistical mechanics: bulk/ensemble properties are calculated from a distribution (Boltzmann or otherwise) under the assumption that the sampled points are representative of the system's phase space. If we fail to sample any system configurations that are energetically low-lying — and therefore non-negligible contributors to the ensemble average — then the computed thermodynamic properties will not mirror the true properties of the system. If, however, a simulation does not get trapped, we are left with a useful result: since the system can explore all of phase space, the distribution of conformations along a simulation trajectory for just one particle will be indistinguishable from the distribution for a solution of many particles at one instant. This is the *ergodic axiom*: all accessible microstates are visited, subject to some p.d.f. that defines the system, in the limit of infinite time/sufficient sampling. Alternatively, the time average of an observable, $\boldsymbol{A}$, for a single particle (denoted $\overline{\boldsymbol{A}}$) equals the ensemble average of that quantity (denoted $\langle \boldsymbol{A} \rangle$) for a macroscopically large set of those particles [85], as expressed below:

$$\overline{\boldsymbol{A}} = \lim_{\tau\to\infty} \frac{1}{\tau}\int_{t=0}^{\tau} \mathrm{d}t\boldsymbol{A}\big(\boldsymbol{p}^N(t), \boldsymbol{r}^N(t)\big)$$
$$\Updownarrow \qquad (4)$$
$$\langle \boldsymbol{A} \rangle = \int \cdots \int \mathrm{d}\boldsymbol{p}^N \mathrm{d}\boldsymbol{r}^N \boldsymbol{A}(\boldsymbol{p}^N, \boldsymbol{r}^N)\boldsymbol{\rho}(\boldsymbol{p}^N, \boldsymbol{r}^N)$$

In these equations, $t$ and $\tau$ indicate time; $\boldsymbol{r}^N$ and $\boldsymbol{p}^N$ are generalized coordinates and momenta of each particle as a function of the $N$ degrees of freedom ($6N$-dimensional integral over all DoF); and $\boldsymbol{\rho}$ denotes the equilibrium phase space probability density function, given, for example, by Eq 2.

## 4. **Molecular dynamics simulations**

The motivation for MD simulations is manifold, and includes studies of protein function (e.g., dynamical basis of allostery [99, 125]), protein malfunction (e.g., effect of point mutations that alter the intrinsic catalytic efficiency of an enzyme in metabolic diseases [126, 127]), the mechanism of protein self-assembly into fibrils and other polymers in neurodegenerative diseases [128], nucleic acid conformational transitions [129], the dynamical basis of specific (and non-specific) protein···nucleic acid recognition [130], the dynamical features of the binding of drug compounds or small-molecule ligands to receptors [131, 16], and other types of molecular recognition events. An overview of the MD method is given in Box 3 and Fig 5.

### 4.1. Why simulation as a route to dynamics?

MD simulations are just that – *simulations* – because many of the timescales relevant to the biological functions of proteins and nucleic acids (Fig 2) are experimentally inaccessible. The functional dynamics of a biopolymer modulate its intra– and inter–molecular interactions and are of great physiological importance. For instance, an enzyme's 'breathing' motions may permit substrates to diffuse into its active site and subsequently re-organize into a productive substrate–enzyme complex [132]. The thermal fluctuations mediating these and other biomolecular recognition events can range from large-scale domain rearrangements and binding/unbinding events to much smaller-scale changes (e.g., redistribution of rotameric states of the conserved side-chains lining an active site). In addition to this example of enzymes, detailed molecular dynamics are what govern the inter-atomic interactions occurring as ligands approach their cognate binding sites, such as in the binding of agonists or antagonists to receptors [133]. There are two key aspects of a molecule's dynamics to consider: The characteristic time– and length–scales that describe the frequency and spatial extent of the motion [7]. As described in §3.3, large-scale motions are intrinsically complex and can occur as combinations of many fundamental modes, harmonic or otherwise; such motional modes are referred to as the *collective modes* that mediate *rare events*. The difficulty of accessing such dynamics via experimental approaches is what motivates modern MD-based simulations.

### 4.2. Overview and justification of the method

By an MD *trajectory* we mean a list of positions and momenta of each particle in a system over time, as the system samples its phase space (Fig 3). The complexity of even a simple biomolecular system – in terms of the number of particles, degrees of freedom, and potential interactions – prevents us from analytically solving for such dynamics using the equations of classical mechanics. Instead, we compute trajectories by approximating the equations of motion via numerical integration: the instantaneous force acting on each particle $i$ is calculated, $\vec{F_i} = -\nabla\mathcal{U}_i$, the forces are used to compute accelerations, and the accelerations are used to update particle velocities and positions. Is this valid? Is it reasonable to perform classical MD simulations (versus quantum dynamics) of protein-sized entities? Does our dynamics method need to treat both the electrons and atomic nuclei?

These questions can be addressed by considering two approximations rooted in the physics of molecular systems: the thermal de Broglie wavelength and the Born-Oppenheimer approximation. The thermal de Broglie wavelength ($\Lambda$) for a particle of mass $m$ is given by $\Lambda = h/\sqrt{2\pi m k_\mathrm{B} T}$, where $h$ is the Planck constant, $k_\mathrm{B}$ the Boltzmann constant, and $T$ the absolute temperature. Of most importance is the value of $\Lambda$ relative to the mean inter-particle separation in the system, $\langle r_{i,j} \rangle$. For length-scales on which $\Lambda \ll \langle r_{i,j} \rangle$, particle interactions can be approximated as classical rather than quantum mechanical [85]. Thus, while the dynamics of light atoms (e.g., mass of hydrogen) at low temperatures ($T \approx 0$) would require quantum mechanical treatment, classical dynamics is a valid approximation for protein-sized entities at typical temperatures of interest in biology ($\approx 300$ K). As for the electronic components of the molecular dynamics, we can neglect these and treat only nuclear motions because of the Born-Oppenheimer approximation [8]. This principle results from the fact that electrons are so much lighter than nuclei that the electron density 'moves' (in response to a force) two orders of magnitude more quickly than do the nuclei. Thus, we can view atoms in a protein as consisting of electron clouds that respond virtually instantaneously to shifts in nuclear positions; more formally, the quantum mechanical wavefunction, which describes the full dynamics of the system, is separable and can be factorized into nuclear and electronic components, given by a pair of Schrödinger equations. In this way, the electronic degrees of freedom are essentially absorbed into the effective interatomic potentials (i.e., force-fields) used in classical MD simulations.

Any MD-based methodology relies on two essential components, one physicochemical (force-fields; §4.3) and one algorithmic (integrators; §4.4). Regardless of the above issue of classical versus quantum dynamics, the core problem in MD — *integrating the equations of motion* — simply requires a set of forces with which to update atomic positions. The algorithm is agnostic about the source of the forces, which can come from *ab initio* quantum mechanical calculations or, as is done in classical MD simulations, by computing force as the gradient of an empirical force-field (below). Note that while the nuclear motions are treated classically, the interatomic forces and electronic structure still can be evaluated quantum mechanically at any desired time-step in the trajectory. Though beyond the scope of this article, hybrid quantum mechanics/molecular mechanics (QM/MM) and *ab initio* MD approaches are essential in order to model processes wherein the electronic structure of a molecule is altered, such as the covalent bond transformations that may occur in enzyme catalysis [8, 134].

### 4.3. Force-fields and the potential energy surface

A force-field (FF) encapsulates all that we believe to be important about the physicochemical properties (§3.2) of the atomic interactions that govern molecular structure & dynamics. As illustrated in Fig 4, the FF expresses molecular interactions quantitatively, using equations, free parameters, and estimates of parameter values. Macromolecular FFs, also known as *potentials*, originated in molecular mechanics efforts of the late 1960s [135]. Those efforts were aimed at calculating primarily structural and stereochemical properties of small organic molecules — conformational strain, geometry optimization, etc. In principle, computing the FF energy as a function of 3D structure, for all possible 3D conformations, would provide the complete potential energy surface of a molecule.

The two defining features of a FF are its general functional form and the precise numerical values it assigns to the constant parameters in its equations. Many FF implementations derive from the following general equation, which gives the potential energy, $\mathcal{U}(\vec{\boldsymbol{r}}_i)$, as a function of position for each atom $i$. In this classic MM approach, covalent interactions are taken as summations over 1–2, 1–3, and 1–4 bonded terms, while non-bonded interactions are modeled pairwise, as sums over Lennard-Jones and Coulombic potentials:

$$\mathcal{U}(\vec{\boldsymbol{r}}_i) = \sum_{bonds} k_i^r (r - r_0)^2 + \sum_{angles} k_i^\theta (\theta - \theta_0)^2 + \sum_{torsions} k_i^\varphi [1 + \cos(n_i\varphi_i - \delta_i)] + \underbrace{\sum_i \sum_{j\neq i} 4\varepsilon_{ij}\left[\left(\frac{\sigma_{ij}}{r_{ij}}\right)^{12} - \left(\frac{\sigma_{ij}}{r_{ij}}\right)^{6}\right]}_{Lennard-Jones\ 12-6\ potential} + \underbrace{\sum_i \sum_{j\neq i} \frac{q_i q_j}{\epsilon r_{ij}}}_{Coulombic\ term} \tag{5}$$

The FF parameters, which may number well into the hundreds, list all the spring constants ($k$), reference bond lengths ($r_0$) and angles ($\theta_0$), torsional angles ($\varphi$), multiplicities ($n$) and phases ($\delta$), Lennard-Jones parameters ($\varepsilon$, $\sigma$), and partial charges ($q$), contained in Eq 5, for all possible types of atoms and pairwise interactions encountered in typical biomolecular systems. While bond lengths and angles are handled in a fairly straightforward and similar manner in different FFs, various biomolecular FFs treat torsional potentials and other terms in subtly different ways. For instance, AMBER and OPLS use specific scaling factors for vdW or electrostatic interactions between 1–4 atoms [136], and some CHARMM FFs employ grid-based energy correction maps ('CMAP') for protein φ/ψ torsional barriers [41, 137]. Regardless of this variation, for all FFs a working set of values (a 'parameter set') is obtained by optimally fitting the parameters, via linear or nonlinear regression, against libraries of target data [41]. These target data originate from two sources, either empirical measurements (e.g., from thermochemistry, such as heats of vaporization, from structural databases, and so on) or ideal values obtained by QM calculations on small model compounds (charge distributions, torsional barrier heights and multiplicities, etc.). The model compounds are small enough for QM calculations at very high levels of theory, and the compounds chemically resemble the constituents of biopolymers – the alanine dipeptide, blocked amino acids, mono- and dinucleotides, etc. For these reasons, the FFs used in MD simulations or docking are said to be *parameterized*, and are described as *empirical force-fields*. Most modern FFs are transferable across related classes of compounds, but make assumptions such as pairwise additivity and the neglect of atomic polarizability (see Box 4 for these terms). Because the FF defines a system's internal energy, the accuracy of a simulation is ultimately limited by that of its FF. Approximations are necessary to make simulations feasible, and the simple functional form of typical FF equations represents a compromise between accuracy and computational tractability. For more infor-mation, lucid accounts of FFs can be found in refs. [41, 138, 42, 43, 139, 140, 44], including reviews of available FFs (AMBER, CHARMM, etc.) and their applicability to various classes of biomolecules. (Virtually all modern FFs are applicable to polypeptides, but some have been more finely tuned than others towards nucleic acids, carbohydrates, or lipids.)

### 4.4. Integrating the equations of motion and computing a trajectory

The *integrator* is the core of any MD-based simulation method (Fig 5). The basic algorithm is the following: given atomic positions and velocities at time $t$, compute the force on each atom from the negative gradient of the energy; from classical mechanics, these forces yield the acceleration of each atom ($\vec{F} = m\vec{a}$) which, in turn, is used to numerically integrate the equations of motion and update the coordinates and velocities, bringing us to time $t + \delta t$. In this way, the system dynamics are propagated from $t \to t + \delta t$ to yield the MD trajectory (in general, $\delta t \approx 1$ fs). Recall that potential energy is determined by the 3D structural coordinates, in conjunction with the FF; kinetic energy is computed from atomic velocities, using Eq 9 in §4.7.2. In principle, each of a system's $N$ atoms could interact with any other atom via bonded or non-bonded interactions, if not at time $t$ then possibly at another time. Therefore, assuming the overall problem can be decomposed into pairwise interactions, and absent any simplifying numerical assumptions or algorithmic tricks, the computational complexity of the core MD calculation scales as $\mathcal{O}(N^2)$; this 'inner loop' over pairwise interactions is the main bottleneck in MD codes, as elaborated below and in refs [141, 104, 81, 142]. In practice, the scaling can be improved to $\mathcal{O}(N \log N)$ via cutoff schemes, particle-mesh Ewald methods, and other approaches described in §4.5.

Because the equations of motion cannot be integrated analytically, many algorithms have been developed for numerical integration by discretizing time ($\delta t$) and applying a finite difference integration scheme [143]; textbooks on differential equations can be consulted for the mathematical bases of these methods (e.g., [144]). MD integrators differ in their balance between numerical efficiency (greater number of simplifying assumptions) and accuracy (fewer assumptions), and the closely related issue of robustness — How sensitive is trajectory stability to timestep $\delta t$? Using a larger $\delta t$ would yield a longer trajectory, but the larger timestep also may render the dynamics unstable, with energies diverging, the protein structure 'exploding', etc. To demystify MD integrators (black box in Fig 5A), the remainder of this section sketches a simple derivation of the 'leapfrog' method (Fig 5C).

To derive the leapfrog integrator, begin by considering the location, given by the position vector $\mathbf{r}$, of a particle at time $t$. Express the position $\mathbf{r}$, velocity $\mathbf{v}$ (first time-derivative of position, also denoted by a single prime $\mathbf{r}'$), acceleration $\mathbf{a}$ (second derivative), and all higher-order derivatives of the particle dynamics ($\mathbf{r}^{(n)}$), as Taylor series expansions in $\delta t$, thereby arriving at the following set of equations:

$$\mathbf{r}(t+\delta t) = \mathbf{r}(t) + \delta t\mathbf{v}(t) + \tfrac{1}{2}\delta t^2\mathbf{a}(t) + \tfrac{1}{3!}\delta t^3\mathbf{r}'''(t) + \cdots \quad (6a)$$

$$\mathbf{v}(t+\delta t) = \mathbf{v}(t) + \delta t\mathbf{a}(t) + \tfrac{1}{2}\delta t^2\mathbf{r}'''(t) + \tfrac{1}{3!}\delta t^3\mathbf{r}^{(4)}(t) + \cdots \quad (6b)$$

$$\mathbf{a}(t+\delta t) = \mathbf{a}(t) + \delta t\mathbf{r}'''(t) + \tfrac{1}{2}\delta t^2\mathbf{r}^{(4)}(t) + \tfrac{1}{3!}\delta t^3\mathbf{r}^{(5)}(t) + \cdots \quad (6c)$$

Relative to the $\mathbf{r}$, $\mathbf{v}$, and $\mathbf{a}$ leading terms, the products with higher-order derivatives can be taken as $\approx 0$ because of the $(\delta t)^n$ coefficients; similarly, equations with third- and higher-order leading terms are not shown above. Truncation of the above series at the third-order derivatives (i.e., all terms higher than acceleration) gives the set of familiar kinematics equations, such as the result that the velocity at time $t + \delta t$ can be computed from the positions at the start and finish of a $[t, t + \delta t]$ time interval: $\mathbf{v}(t) = (\mathbf{r}(t + \delta t) - \mathbf{r}(t))/\delta t$ (indeed, this is the definition of the derivative, in the limit as $\delta t \to 0$). Now, recall from calculus that the mean value theorem ensures that for any differentiable function $f$ there exists some point $b$, in a closed interval $[a, c]$ from the domain of $f$, at which the derivative $f'$ equals the average value of the slope across the entire interval (note that this mean value, the slope at point $b$, corresponds to the slope of the secant line passing through $a$ and $c$). In other words, there exists some point $b \in [a, c]$ such that $f'(b) = (f(c) - f(a))/(c - a)$ holds true. Application of the mean value theorem to the third-order truncated form of the foregoing Taylor expansions of the position $\mathbf{r}(t)$ and velocity $\mathbf{v}(t)$ establishes the two halves of the leap-frog integrator equations (Fig 5C). Note from the leap-frog equations that a slight computational inconvenience of this algorithm is that the position and velocity updates are offset by half a step, $\delta t/2$ (Fig 5C); this inconvenience becomes asymptotically negligible as we approach the number of integration steps typical in biomolecular simulation (for $\delta t = 1$ fs, a 50-ns trajectory takes $5 \times 10^7$ integration steps). The 'order' of an integrator, denoted using big-$\mathcal{O}$ notation as $\mathcal{O}(N)$ for order $N$, is the highest-order term in the above series expansion which is included in the calculation (i.e., it denotes the level of approximation). Because leap-frog equations neglect derivatives of order three and higher, this integrator is said to be a *second order* method [104]. Many further issues concern the application of integrators in MD, such as trajectory stability (robustness of the integrator), multiple time-stepping schemes, time reversibility, suitability of the integrator equations for simulating dynamics in various ensembles and constraint schemes (freezing-out fast motional modes, such as sub-fs vibration of X–H bonds, enables larger timesteps). These and related topics are discussed in many simulation and modeling texts, including those by Leach [8], Schlick [10], Haile [104], and Allen & Tildesley [141]. Also, Frenkel [145] and others (e.g., [146]) have recently presented some of the potential pitfalls inherent to simulation studies.

### 4.5. Optimizing the integrator

The evaluation of bonded terms in the system Hamiltonian is straightforward. The integrator maintains a list of all bonded interactions in the system, known from the covalent structure, and, at each timestep, evaluates these energies. Bonded interactions correspond to bond distances, angles, and dihedrals, and therefore are also known as 1–2, 1–3, and 1–4 terms. Since the number of bonded interactions grows as $\mathcal{O}(N)$ for $N$ particles, the evaluation of bonded energy terms is also $\mathcal{O}(N)$. (In classical MD, bonds are not altered and the electronic structure of the molecule is preserved.) Non-bonded forces are more expensive to calculate because there are many more of them: Both vdW and electrostatic interactions occur between all pairs of atoms, and for $N$ atoms there are $N(N-1)/2$ such pairs. Therefore, naïve algorithms for non-bonded forces would scale as $\mathcal{O}(N^2)$ [81].

To optimize the non-bonded vdW force evaluation, a cutoff radius is chosen, typically near 10 Å. All pairwise interactions exceeding the cutoff radius are assumed to be negligible. A side-effect of this method is a discontinuity in energy as atoms cross the cutoff distance, which creates an infinite spike in the energy gradient and therefore an infinite force; such a scenario leads to system instability [81] and lack of energy conservation. Smoothing functions alleviate this problem by removing the discontinuity; specifically, forces are evaluated using the vdW potential up to a 'switching distance' (often ≈9 Å), and the vdW potential is smoothly decreased to a value of zero at the ≈10 Å cutoff (see, e.g., the NAMD user guide for a technical discussion [147]). Though this algorithm makes $\mathcal{O}(N)$ energy evaluations, it still must make $\mathcal{O}(N^2)$ distance evaluations — the distance to each other atom must be checked against the cutoff distance. To avoid this bottleneck, nonbonded 'pair-lists' are used to track which atoms may be within the cutoff. A pair-list distance (≈12 Å) is chosen, and each atom keeps a record of which other atoms were within this distance. Then, during the evaluation of vdW energies, only atoms in the pair-list are considered. After multiple time-steps, pair-lists are updated to account for atoms that may have moved in/out of the distance limit; these updates must occur often enough that no atom moves from outside of the pair-list distance to inside the cutoff distance before a regeneration cycle. Though regenerating the pair lists is $\mathcal{O}(N^2)$, it occurs only infrequently, and can be further reduced to $\mathcal{O}(N)$ using cells, as described below [142].

Electrostatic interactions decay less rapidly than vdW interactions, so using a simple cutoff scheme to define the set of necessary force calculations would require a very large cutoff. Instead, the electrostatic interaction for the simulation system and all its periodic images (an infinite crystal) is generally treated using the particle-mesh Ewald (PME) approach, which decomposes the electrostatic energy into two parts: a short-range component that is evaluated with high accuracy, and a long-range component that is approximated via discretization of charges on a grid and calculations of reciprocal-space structure factors using this mesh. The short-range component is evaluated in real (or 'direct') space in much the same way as the vdW energy described above, and therefore goes as $\mathcal{O}(N)$. The long-range component is generally evaluated using Ewald sums in Fourier (or 'reciprocal') space, using the fast Fourier transform (FFT) for calculation of structure factors and spline-based interpolation of the reciprocal-space sum [148]. FFT calculations scale as $\mathcal{O}(N \log N)$, so the overall scaling of the popular PME algorithm is also $\mathcal{O}(N \log N)$ [149]. A detailed description of the PME method is beyond the scope of this work; further information can be found in refs [148, 8].

Computationally expensive MD simulations are frequently performed on distributed-memory supercomputers. In a distributed-memory supercomputer, a set of independent computers are connected by a high-speed network, and each computer is responsible for simulating some regions of the overall system. Each region periodically informs its neighbors about the movement of its atoms. If the regions are larger (in every direction) than the cutoff distance, then each region needs to communicate only with its 26 neighbors [142]. Since most communication occurs between neighboring regions (some global communication is necessary for electrostatics and monitoring), atomistic MD codes such as NAMD [142] can scale to hundreds of thousands of processors with systems consisting of tens of millions of atoms [150]. Other available MD codes for biomolecular systems include popu-lar, long-standing software suites such as AMBER [136], CHARMM [151], GROMACS [152] and LAMMPS [153], newer packages such as DESMOND [154] and TINKER [155], and a host of other programs [156].

### 4.6. Some practicalities: From theory to practice

Beyond the integrator, many practical questions must be considered in preparing to simulate. The first stage of any simulation (*system setup*) is to prepare the molecular system, which includes the biomolecular solute and any solvent, ion, ligand, etc. molecules. The components of a 'simulation system' are defined in Fig 5A and its caption. Simulations are typically performed in the $NVT$ or $NPT$ ensemble in order to mimic experimental conditions as closely as possible. Periodic boundary conditions (Fig 5B) help avoid surface effects (i.e., mimic bulk solvent), though of course real solution-phase systems are not loosely packed crystals; as mentioned above, long-range electrostatics are handled in periodic systems via Ewald sums [157]. Suitable system setup also requires one to consider *(i)* many chemical details (protonation states of ionizable residues, ionic strength, etc.); *(ii)* which force-fields to use (plural, if comparative analyses are being undertaken such as in [158] or [140]); *(iii)* addition of solvent molecules, and choice of water model [159]; *(iv)* decisions regarding non-bonded cutoff distances and switching functions; and, finally, *(v)* possible preliminary stages of energy minimization to relax the starting structure by relieving high-energy inter-atomic contacts (up to this point, the 3D structure of the biomolecule has not 'seen' the potential energy surface defined by the chosen FF). Once these setup stages are complete, the MD system is subject to a brief *heating and equilibration phase* (often ≈ 1-5 ns), followed by a *production phase* of free dynamics, during which time atomic coordinates and velocities are written to disk every few ps. Common practices in setting-up simulations are available in the literature for generic biomolecular systems [160, 124], nucleic acids [161], and membrane proteins [162]. With modern computing power, production-length simulations are often ≈50-100 ns, and the longest are on the order of milliseconds [163, 69]. The exact duration depends on a balance of system size, compute resources, user patience, the timescale of the biochemical questions motivating the simulation (Fig 2) and, ultimately, how much sampling is required to achieve converged structural or dynamical properties.

Seemingly abstract simulation concepts can have very tangible consequences for how one proceeds in performing an MD calculation. These consequences are of practical concern to the user of a software package, and can be transparently understood in terms of basic principles. For example, consider the role of cutoffs in all-atom MD simulations. One is generally interested in the dynamical properties of a protein in bulk solution, not isolated in a nm-sized droplet of water, which would place the protein at a water/vacuum interface. To avoid potentially artefactual surface effects, a protein is simulated under periodic boundary conditions (PBC; Fig 5). The PBC geometry is essentially a highly solvated crystal, packed loosely enough so that solute···solute interactions across cells are negligible (e.g., the DNA in Fig 5B and its periodic images). An inherent geometric property of PBCs is that the so-called *minimum image convention* must be applied in order to avoid over-counting particle interactions (see the central cell and colored balls in the upper-right of Fig 5A) [141, 8]. This, in turn, demands the use of distance cutoffs for evaluation of non-bonded forces and long-range electrostatics. As the final step in this chain of implications, note that the cutoff distance ($r_{cut}$) cannot exceed half the cell edge (for simplicity consider a cubic unit cell), lest pairwise interactions be double-counted. The above line of reasoning is motivated by conceptual factors, such as the desire to simulate a biomolecule in bulk solution rather than at an air/water interface. A second, practical reason for using cutoffs, to lessen computational costs, was described in §4.5. A cutoff distance $r_{cut}$ ≈ 10-11 Å is generally used in biomolecular simulations, as a compromise between accurate evaluation of enough nonbonded forces (necessary for trajectory stability, energy conservation, etc.), versus excessive computational cost (the number of pairwise nonbonded forces to be evaluated scales as $r_{cut}^3$ [142]).

### 4.7. Interpreting and assessing results

#### 4.7.1. Trajectory analysis via RMSD

The raw data from a simulation is a list of coordinates $\mathbf{r}_i(t)$ and velocities $\mathbf{v}_i(t)$ as a function of time ($t$) for each atom $i$. That is the *trajectory*. These data can be structured as a pair of two-dimensional arrays, $\mathbf{R}$ and $\mathbf{V}$. Matrix $\mathbf{R}$ is built from the $3N$ Cartesian coordinates $(x, y, z)$ for the $N$ atoms and, in the other dimension, is index by the simulation time $t$ (and similarly for the velocities, $\mathbf{V}$). Equivalently, a column vector of $\mathbf{R}$ gives all the coordinates for all atoms at one time-step, while a row vector gives the time series of a particular coordinate of one atom. Coordinates are typically written every ≈1-2 ps, meaning that millions of snapshots are created in a µsec-scale simulation. What knowledge can be extracted from such dense data?

Simulation analyses can range from routine and straightforward to highly sophisticated, and can be either highly generic or more specialized to the type of system/question at hand. An example of a generic type of trajectory analysis, applicable to any system, is computation of the root-mean-square deviation (RMSD) of coordinates over time. Though the RMSD is not always an ideal metric for assessing equilibration and structural stability [164], an RMSD analysis is performed early on (within the first few ns) in virtually all atomistic simulation studies [24, 160]. The RMSD for two coordinate sets, $\mathbf{s}_x$ and $\mathbf{s}_y$, is readily defined as:

$$\mathrm{RMSD}(\mathbf{s}_x, \mathbf{s}_y) = \sqrt{\frac{\sum_{i=1}^{N} \left\| \vec{r}_{\mathbf{s}_x,i} - \vec{r}_{\mathbf{s}_y,i} \right\|^2}{N}}. \qquad (7a)$$

In this formula, $\vec{r}$ is the vector of position coordinates for each of $N$ atoms, with each atom pair indexed by $i$; $\mathbf{s}_x$ and $\mathbf{s}_y$ can, for example, be two frames in a trajectory (i.e., columns $x$ and $y$ of $\mathbf{R}$). Closely related, the root-mean-square fluctuation (RMSF) of atom $i$, in a structure evolving over time, $\mathbf{s}(t)$, can be formulated as:

$$\mathrm{RMSF}(\mathbf{s}(t), i) = \sqrt{\frac{\sum_{n=0}^{f} \left\| \vec{r}_{\mathbf{s},i}(t_n) - \langle \vec{r}_{\mathbf{s},i} \rangle \right\|^2}{f+1}}, \qquad (7b)$$

where now the summation is performed over all time-steps of interest (from $n = 0 \rightarrow f$), and $\langle \vec{r}_{\mathbf{s},i} \rangle$ and $\vec{r}_{\mathbf{s},i}(t_n)$ are the time-averaged and instantaneous ($t_n$) coordinates of atom $i$, respectively. Slightly more sophisticated, one can compute two-dimensional matrices of pairwise RMSDs, thereby avoiding the issue of precisely which 3D structural snapshot should be taken as the reference point for the calculation (the starting structure?, after 1-ns of equilibration?, an averaged structure?, etc. [129]).

#### 4.7.2. Principal component analysis and related approaches

As an example of a more sophisticated analysis approach, principal component analysis (PCA) can be used to calculate the directions and amplitudes of greatest motion along a simulation trajectory. PCA is a linear algebraic method of 'dimensionality reduction', meaning it can map datasets of high dimensionality — e.g., the vast vector space ($\mathbb{V}$) spanned by the $3N$ coordinates of a simulation system, sampled across millions of timesteps (the matrix $\mathbf{R}$ in §4.7.1) — into a new vector space ($\mathbb{V}'$), defined by an alternative basis set. The key feature of the PCA approach is that this new, alternative basis set spans the bulk of the variation (literally, the statistical variance) that occurs in the original high-dimensional data, and it does so in a more informative manner than does the original/naive basis set: We obtain a rank-ordering of the fraction of variance that is accounted for along each new basis vector, and the major directions of motion can be expressed as simple linear combinations of the new basis vectors (also known as principal component vectors, as described below). A major strength of PCA is that it is a *non-parametric* method for analyzing high-dimensional datasets, such as the many frames comprising an MD trajectory. PCA is free of heuristics, assumptions about dynamical modes, etc., and the PCA algorithm takes trajectory data as its only input. A fundamental limitation of PCA is that the $\mathbb{V} \rightarrow \mathbb{V}'$ mapping alluded to above is a linear transformation; therefore, subtle non-linear correlations will be missed, such as correlated motion along circular paths (see the 'Ferris wheel' example in [165]). PCA captures only that underlying structure of the data that is expressible as linear correlations.

Useful introductions to PCA are available, from both general (e.g., [166, 165]) and MD-specific (e.g., [167]) perspectives. In brief, consider a trajectory comprised of $m$ frames, for a simulation system of $N$ atoms. Begin by removing the six rigid-body translational and rotational degrees of freedom of the molecule via least-squares structural superimposition of each frame to a reference (e.g., the initial structure). Then, construct a $3N \times m$ matrix, $\mathbf{R}$, from the $3N$ Cartesian coordinates at frames $1, 2, \ldots, m$. In this matrix, column $j$ is the vector of all atomic coordinates at frame $j$. PCA is then achieved by *(i)* using $\mathbf{R}$ to construct the variance-covariance matrix, $\mathbf{C}$, of 3D coordinate displacements $\vec{r}$ (versus trajectory-averaged mean coordinates, $\langle \vec{r} \rangle$), and then *(ii)* diagonalizing $\mathbf{C}$ to obtain the *principal components* of the motion, denoted $\vec{p}_i$, as projections onto the eigenvectors of this covariance matrix [168, 169]. These two steps correspond to the following pair of equations:

$$\mathbf{C} = \langle \mathbf{R}\mathbf{R}^{\mathrm{T}} \rangle = \langle (\vec{r}(t) - \langle \vec{r} \rangle)(\vec{r}(t) - \langle \vec{r} \rangle)^{\mathrm{T}} \rangle, \qquad (8a)$$

$$\mathbf{C} = \mathbf{Y}\boldsymbol{\Lambda}\mathbf{Y}^{\mathrm{T}}, \qquad (8b)$$

where $\mathbf{Y}$ is the orthogonal transformation that we seek to discover to diagonalize $\mathbf{C}$, $\boldsymbol{\Lambda}$ is a diagonal matrix containing the corresponding eigenvalues ($\lambda$'s), and a superscript 'T' denotes the transpose. Note that $\mathbf{C}$ is a symmetric $3N \times 3N$ matrix, from which the linear cross-correlation matrix is obtained simply by normalizing each element $c_{i,j}$ by the factor $(c_{i,i}c_{j,j})^{1/2}$; viewed this way, the diagonal elements of $\mathbf{C}$ are the mean-square atomic fluctuations, $\langle|\Delta\vec{r}_i|^2\rangle$, that appear in §4.7.3 below. The columns of $\mathbf{Y}$ are the eigenvectors of $\mathbf{C}$. The original trajectory coordinates, $\mathbf{R}$, can be projected onto these eigenvectors, $\vec{u}_i$, in order to visualize the motion along each of those directions; doing so gives the corresponding $\vec{p}_i$ principal components. Notably, the eigenvectors $\vec{u}_i$ are sorted in decreasing order of their corresponding eigenvalues, $\lambda_i$. Thus, eigenvector $\vec{u}_1$ is the direction along which the greatest motion occurs – i.e., the direction that accounts for the largest fraction of variance in atomic positions across the dataset. The corresponding eigenvalues give the statistical variance along each mode – i.e., the amplitude of motion, measured as mean-square displacements. For proteins, an empirical finding is that the first several $\vec{u}_i$'s account for much of the variance in atomic positions, at least on relatively short timescales where the assumption of linearity is unlikely to break down; for this reason, PCA is also known as 'essential dynamics' [168]. High-amplitude vectors, which correspond to low-frequency modes in the harmonic approximation, are often taken as being functionally important dynamical modes; for instance, a 'hinge' between two protein secondary structural elements, about a specific direction given by principal component $\vec{u}_i$, and with a particular magnitude ($\lambda_i$), may elucidate the dynamical basis for 'gating' of an active site. As a further step, one may pursue clustering of protein conformers in the reduced dimensionality space of the first few principal components (the $\mathbb{V}'$ space, above), rather than in the original Cartesian basis; such approaches can be used to compare the 'essential subspaces' of the dynamics of different proteins, assess trajectory equilibration, etc. Finally, note that PCA is closely related to other eigenvalue decomposition approaches, such as normal mode analysis (NMA) and quasi-harmonic analysis (QHA). For example, mass-weighting the terms in the covariance matrix gives the QHA approach which, in turn, can be used to estimate the conformational entropy from an MD trajectory [170-172].

In addition to a PCA decomposition of the trajectory, other quantities can be computed by relying on statistical mechanics as the link between raw trajectories (dynamics) and bulk thermodynamic observables. For example, one can compute the velocity autocorrelation function from a trajectory as an estimate of the diffusion coefficient [141]; similarly, other trajectory-derived correlation functions can be calculated and compared to experimentally characterized transport coefficients. As another example of the experiment ↔ *simulation* ↔ theory link, the radial distribution function (RDF) is a versatile theoretical concept that can be computed from trajectories and used in connection with both theory and experiment. As the name implies, an RDF gives the distribution of particles, or *number density*, in a simulation system as a function of radial distance (i.e., isotropically averaged) from a reference particle, averaged again over all relevant reference particles. This equilibrium quantity also can be viewed as the distribution of all distances between all pairs of particles (a spatial pair correlation function), and it is therefore deeply related to the time-averaged structure of a system of particles. In this way, the RDF directly links to experimentally measurable quantities that report on inter-particle separations, such as solution scattering profiles obtained by small-angle X-ray scattering (SAXS) [173, 174]. An MD trajectory provides all coordinates (structures) at every time-point of interest, meaning one can use a trajectory to compute any desired RDF [104] – between all oxygen atoms in water, between a particular set of ions and a particular base in RNA [175], etc. – for joint analysis with experimental scattering data. That is the *experiment* ↔ *simulation* link. As an example of the other direction (*simulation* ↔ *theory*), the RDF is intimately related to the statistical mechanical *potential of mean force* (PMF; [85]), and to the use of the PMF concept to justify the derivation of pairwise statistical (*knowledge-based*) potentials from databases of known 3D structures [176, 177]. Thus, simulation-derived RDFs can also facilitate the testing of theoretical models and approaches.

#### 4.7.3. Reliability, validation, and relative strengths of the simulation approach

Simulations can be viewed as more predictive than conclusive. This is true of any purely computational approach and, indeed, any method taken in isolation (experimental or computational). What simulations lack in certainty, versus a set of carefully-controlled biochemical experiments, they make up for by being the only widely available approach that can provide high resolution information about the dynamics of virtually any biomolecular system, in both space (atomic-resolution) and time (sub-ps time resolution). The 'validity' of a given MD trajectory partly depends on the exact biological question being considered (was the system simulated long enough?), as well as a host of potential technical concerns. These technical issues are numerous and are often system-specific; the remainder of this subsection is limited to a few illustrative points.

Catastrophic errors often manifest themselves early in a trajectory, and often can be readily identified. For instance, the PME approach to long-range electrostatics can be sensitive to electroneutrality of the simulation system: if the simulation cell contains excess electric charge because counter-ions were not added, then lattice sums will diverge to infinity. In practice, whether or not this problem occurs depends on the capabilities of the MD software and its default configuration settings. For instance, non-neutral cells are auto-detected by many MD codes and a uniform 'neutralizing plasma' is applied as another term in the Ewald sum; inclusion/exclusion of this term is akin to the crystallographic $\vec{F}_{000}$ structure factor, the amplitude of which is the number of electrons per unit cell, but which is an arbitrary additive constant in typical electron density map calculations. Errors in constructing a PBC cell can result in atoms unfavorably interacting with other image atoms, yielding energy divergence and trajectory instability. Finally, simulations are also susceptible to less severe (but also more subtle) errors, such as the possibility of periodicity-induced artefacts for the PBC simulations that are customary in biomolecular MD [178].

Efforts to ensure a reliable, or at least stable, trajectory must be made in the earliest stages of system selection and preparation (protonation, addition of ions, solvation, etc.), before the lengthy production phase commences [24, 160, 124]. Successful equilibration is vital, at least to the extent possible [146], and can be judged in terms of both structural stability and conservation of thermodynamic quantities. Structural stabilization of a trajectory can be assessed by monitoring properties such as secondary structural content, by visual inspection in a molecular graphics suite such as VMD [179], and by plotting quantities such as the radius of gyration or RMSD to see that the system has not unfolded or dissociated (in the case of a supramolecular assembly). For thermodynamic equilibration, those bulk properties that are expected to be conserved for the particular ensemble being used should reach stable values, generally within the first few hundred ps of simulation; bulk quantities will fluctuate, but should show no systematic drift. For instance, in addition to conservation of total system energy, one would expect temperature stability for simulations in an isothermal ensemble such as $NPT$. In practice, temperature can be monitored via an instantaneous 'kinetic temperature', $T_k$. This quantity can be computed at timestep $t$ from a trajectory's atomic velocities by using the equipartition principle and the definition of kinetic energy in terms of particle velocities [141]:

$$T_k(t) = \frac{1}{N_f k_\mathrm{B}} \sum_{i=1}^{N} \frac{|\vec{p_i}(t)|}{m_i}, \qquad (9)$$

where $i$ indexes all $N$ particles of momentum $\vec{p}$ and mass $m$, and the other symbols are as used above. The $N_f$ in the denominator of the prefactor is the number of degrees of freedom. This term may equal $3N$ for the components of velocity for a monoatomic particle in three dimensions or, for example, $3N - N_c$ if $N_c$ internal constraints are applied; the exact details depend on the exact dynamical system and simulation protocol. Averaging over many MD time-steps yields $T = \langle T_k(t) \rangle$ as the thermodynamic temperature.

The question of sufficient sampling – how long to run a simulation – is difficult, as it depends on balancing computational cost against the exact meaning of 'sufficient'. As noted above, meaningful precision can be attained with a ten-fold excess of data [104]; however, one may be unsure as to the characteristic timescale for a process of interest (e.g., a conformational transition). Assessment of simulation *accuracy* is yet more difficult, largely due to the limited experimental options for cross-validation. As an example of the type of data that may be used for cross-validation, trajectory-derived root-mean-square fluctuations (RMSF) for each residue in a protein can be compared to patterns of variability from NMR order parameters ($S^2$) [180] or the $B$-factors obtained from refinement against X-ray diffraction data [181, 182]. The $B$-factor, also known as the Debye-Waller factor [183], quantifies the attenuation of X-ray scattering intensity ($I(\vec{h})$) for each peak in a diffraction pattern. Expressed in reciprocal (diffraction) space, with $\vec{h}$ denoting the vector of Miller indices $h, k, l$ for each Bragg reflection, we have:

$$I_{exp}(\vec{h}) = I_0(\vec{h}) e^{-2B(\sin^2\theta/\lambda^2)}. \qquad (10)$$

In this equation, $I_{exp}$ is the measured experimental intensity, $I_0$ is that for the ideal (frozen) lattice with no thermal vibration, $B$ is the overall temperature factor, and the $\sin^2\theta/\lambda^2$ term captures the standard decrease in the magnitude of atomic form factors with increasing Bragg angle ($\theta$) for a given X-ray wavelength (λ). Alternatively, individual $B$-factors can be expressed in terms of individual/atomic motion, in real space, as follows:

$$B_i = \tfrac{8}{3}\pi^2 \langle |\Delta\vec{r_i}|^2 \rangle, \qquad (11)$$

where $\langle \cdots \rangle$ denotes the ensemble average and $\langle |\Delta \vec{r_i}|^2 \rangle$ is the mean square coordinate displacement of atom $i$ about its equilibrium position. These equations, which take $B$-factors as scalars, assume isotropic atomic displacements; given sufficiently high-resolution diffraction data, full anisotropic $B$-factor tensors can be used to better resolve the atomic displacements [183]. Equations 10 and 11 link an experimental observable, namely X-ray reflection intensities, and a simulation-derived quantity, the RMSF along the trajectory (Eq 7b in §4.7.1). However, in attempting to validate a simulation by corroborating MD-derived RMSFs to patterns of variation in crystallographic $B$-factors (high $B$-factors in loops, active site residues, etc.), one should note two issues: *(i)* the typical $B$-factor refinement approach assumes isotropic and harmonic thermal motion, and *(ii)* the $B$-factor values generally computed in macromolecular X-ray refinement implicitly include a host of additional, non-dynamical effects. Issue *(ii)* is important because the mean atomic displacement of a specific residue in a macromolecular crystal arises from the authentic intramolecular dynamics of that residue, but also includes effects of static disorder and microscopic heterogeneity (slight conformational variability in each unit cell), lattice imperfections and vibrations, and so on. Because both static and dynamic phenomena contribute to the attenuation of X-ray reflection intensities, care must be taken when interpreting $B$-factors in terms of specific dynamical processes.

Beyond the above issues, two main factors limit the precision and accuracy of simulations. Firstly, the approximations inherent in force-fields, and the MM approach itself, restrict the accuracy of trajectory-derived values, as alluded to in §4.2–3. Secondly, the necessarily limited sampling means that trajectory-derived numerical averages may be insufficiently converged, there could be many conformational transitions that occur in nature but go unobserved in a limited-length trajectory, and so on. These two limitations – force-fields and statistical sampling – have motivated many areas of contemporary MD research. For instance, much recent work has been devoted to creating polarizable force-fields [184, 185], more efficient *ab initio* and hybrid QM/MM approaches [134], enhanced sampling techniques [71], and so on. A thorough discussion of the relative merits of various MD-based approaches can be found in [47].

### 4.8. Simulations in structural biology

In addition to the utility of simulations in analyzing biomolecular dynamics and function (e.g., allostery), MD-based methods are used in biology to *determine* the 3D structures that serve as starting points for such analyses. Perhaps nowhere has the practical impact of MD been greater than in experimental structural biology, which is largely concerned with determining structures via X-ray crystallography or NMR spectroscopy. To illustrate the power of leveraging MD with experiment, consider the role of simulating annealing refinement. MD-based simulated annealing is generally used in the refinement of both crystallographic models [186] and in NMR structure determination [187]. Simulated annealing refinement works by using MD as a conformational search tool: An artificial energy landscape is constructed by adding a fictitious energy term to the FF, to penalize discrepancies with diffraction data. Using MD, this landscape is initially sampled at exceedingly high (physically unrealistic) temperatures, thereby providing the system — in this case, the trial 3D structural model — with enough thermal energy to cross local energy barriers. Several stages of short dynamics runs are performed, with the system temperature lowered at each stage according to a prescribed cooling schedule. The power of this approach is that it generates successively better (lower energy) structures as the simulation stages proceed at sequentially lower temperatures, thereby refining the 3D structure. Further details of this MD approach and its utility in crystallography have been reviewed [75].

This fruitful application of simulating annealing to structural biology illustrates a general principle: Because of their generality, simulation-based approaches offer flexible frameworks for *handling* experimental data (to get to a 3D structure), *integrating* various types of data, and then *extracting knowledge* that is inaccessible from such data alone (e.g., dynamics of the 3D structure). An example of the synergistic application of computation and experiment is the determination of a structural model for the nuclear pore complex (NPC), an ≈ 50-100 MDa assembly of hundreds of proteins and lipids (reviewed in [1]). In the NPC work, many lines of experimental data were taken as distance restraints and cast as energy terms in a molecular mechanics framework [188]. This approach enabled the application of energy minimization and simulated annealing routines to obtain a collection of structures most compatible with the combined set of experimental data (or least incompatible, in the sense of a cost function). The success of the approach hinged on two facts: *(i)* electron microscopy, chemical cross-linking, mass spectrometry, and virtually any other source of low-resolution data facilitates structure determination [1] by constraining the allowed 3D structures, and *(ii)* computational methods, such as the simulation-based methods of this text, provide a way to sample the space of possible solutions, via rapid generation and evaluation of trial struc-tures. Thus, computational methods provide a natural framework for the development and implementation of 'hybrid' approaches for difficult/low-resolution structure determination.

## 5. **Computational docking as a means to explore molecular interactions**

Because of the pivotal roles of molecular dynamics and interactions *in vivo*, many computational approaches have arisen to model and elucidate these interactions *in silico*. The many different types of molecules (proteins, nucleic acids, small molecules, etc.) found inside even the simplest of cells means that an even far greater number of conceivable types of interactions can occur in cellular physiology [5]. Such interactions are often pairwise, A•B, where if A = B (and the constituents are protein) the interaction is termed *homotypic* (e.g., homo-oligomers such as an ATPase), whereas for A ≠ B the interaction is called *heterotypic* (e.g., a hetero-dimer protein, oxygen bound to hemoglobin). Most generally, the interaction partners A and B may be protein, nucleic acid, carbohydrate, lipid, or any of a number of other small molecules and ions (the heme ring in hemoglobin, ATP-binding sites, etc.). The binary A•B complex may be short– or long–lived with respect to the lifetime of the cell, and the A•B association may be thermodynamically quite stable (e.g., cytoskeletal polymers [189]) or only marginally so (e.g., entrapment of a polypeptide in the GroEL cage for folding and release [190]). Finally, in addition to binary interactions, ternary and higher-order contacts can occur, giving rise to intricate homo- or hetero-oligomeric complexes and, in some instances, open-ended polymeric structures such as the cytoskeletal 'scaffolding' proteins.

### 5.1. Physical chemistry of molecular associations

Apart from the role of macromolecular crowding [2] in promoting interactions between any two random molecules A and B, note that a specific A•B complex will form only once the entities A and B are within suitable distance for energetically favorable inter-atomic interactions to occur, denoted by A···B. What is meant by 'suitable distance'? Recall from §2 and §3 that non-covalent forces originate in the laws of physics, and are of only a few fundamental varieties: relatively long-range electrostatics ($\mathcal{U}_{elec} \sim 1/r$), shorter-range hydrogen-bonding interactions (fundamentally electrostatic, requires chemically complementary donor and acceptor), and even shorter range vdW interactions (features attractive and repulsive components). In addition, solvation and other entropic effects play a major role in molecular interactions [191-193]; these effects include the entropy-driven free energy changes due to solvent reorganization and differential exposure of hydrophobic patches near the A•B interface. In computing the affinity of an A···B contact, note that a possibly delicate balance of entropic effects is at play: Taking A and B as rigid bodies, six rotational and translational degrees of freedom are lost upon formation of the complex ($\Delta S^o_{\mathrm{A \bullet B}} < 0$, disfavoring association), while the entropy change of solvent molecules liberated from the A•B interface ($\Delta S^o_{solv}$) will favor association. In one common approach, the magnitude of $\Delta S^o_{solv}$ is taken to be proportional to the solvent-accessible surface area that becomes occluded in the A•B interface [194]. Though far from straightforward, properly accounting for these subtle entropic effects is necessary for accurate calculations of ligand-binding free energies [89, 107, 45].

Formation of an initial A···B 'encounter complex' occurs via the diffusional association of A and B, followed by possible smaller–scale intermolecular interactions and intra-molecular rearrangements (*induced fit*) that finely tune the stability of the complex. An alternative model of ligand-binding mechanics is *conformational selection* [195-197], wherein the ligand B binds favorably to a particular subset of conformers of A, 'selected' from the full ensemble of thermally-accessible states of A under the given conditions. Features of both the induced fit and conformational selection models are likely to occur in many ligand-binding reactions [198]. For both models, the molecular interactions are precisely the sorts of nonbonded forces listed above and in §3.2, and are what one attempts to correctly capture for accurate protein–ligand docking. Hydrodynamics and its associated methods, such as Brownian dynamics simulations, provides the theoretical and computational framework for studies of diffusional association and dissociation of A···B over cellular length-scales (≈ tens of nm) and timescales (μs → ms) [199, 119]. These length and time regimes generally exceed what is possible, in terms of both algorithmic frameworks and computational resources, for studying the fine-grained (atomic-level) details of A···B interactions — for instance, elucidating specific hydrogen bonds between a patch of conserved amino acids on A and a structurally complementary region of B, the *open* ↔ *close* dynamics of a hydrophobic trench on the surface of A, etc. These two problems of *(i)* long-distance, long-time diffusional association of A and B, and *(ii)* short-distance, short-time details of interactions between A and B (and molecular dynamics of the resultant A•B complex) are essentially handled as separate issues in current computational studies, rather than treated in an integrated manner. The remainder of this section focuses on methods to study P···L and P···P interactions, where one entity is pro-tein ($P$; also termed the *receptor*) and the other component may be a small-molecule compound known as the *ligand* ($L$).

In principle, the computational approaches developed to treat receptor···ligand interactions can be generally applied to any A···B system, be it protein···protein, protein···nucleic acid, nucleic acid···ligand, etc. In practice, the variations between these types of interactions enable different sets of approximations and methods to be applied to each. As with force-fields and molecular mechanics, the calculations are numerically intensive, and simplifying estimations are necessary to render the calculations feasible. For instance, crude treatment of magnesium ions is unlikely to degrade the overall results of a protein–ligand docking pipeline, as magnesium plays a relatively rare role in mediating such interactions in proteins; however, deficiencies in modeling $Mg^{2+}$ would adversely affect RNA–ligand docking, as many such interactions are magnesium-mediated (polyvalent ions are a weakness in typical all-atom classical MD simulations with non-polarizable FFs [200]). Because protein···ligand and protein···protein docking have been the most thoroughly studied, the remainder of this section focuses on these two types of molecular interactions.

## 5.2. Protein–ligand docking

### 5.2.1. General goals

Unlike the usage of MD to study the conformational dynamics of a protein, the general goal of most protein–ligand docking efforts is not to simulate the binding process as it occurs in nature (a notable exception is ref [16]). Rather, the aim is to predict and characterize possible molecular complexes in terms of the 3D structure of the ligand-binding site and the ligand itself (the *pose*), and possibly the ligand-binding energetics as well [89]. (The standard state Gibbs free energy of binding is a measure of equilibrium binding affinity via the usual relationship $\Delta G^o_{bind} = -RT \ln K_D$.) After an initial round of docking studies, one may wish to carefully dissect the enthalpic and entropic components of binding, $\Delta G^o_{bind} = \Delta H^o - T\Delta S^o$, in order to use such information to guide and refine the ligand design process. For instance, decreasing the number of rotatable (single) bonds in a candidate inhibitor compound will reduce its entropy loss upon binding, thus enhancing the overall binding affinity (all other things being equal) [201].

### 5.2.2. More specific goals

In planning a docking study, the precise objectives must be carefully considered, as these goals dictate the allowable approximations in the scoring method and the necessary amount of sampling. Three scenarios can be envisaged. (1) Is the goal to exhaustively characterize the binding of a single compound L across the surface of a protein $\mathcal{P}$? If so, then extensive sampling across the entire protein surface must be performed ('blind-docking' assumes no knowledge of the location of potential ligand-binding sites), with moderate approximations necessary for the scoring function [202]. (2) Is the goal *virtual screening* (Box 5) of large databases of compounds against protein $\mathcal{P}$? If so, then the degree of sampling will be necessarily quite limited and more aggressive approaches for rapidly generating trial configurations, such as genetic algorithms or Monte Carlo, must be employed, versus more physically realistic (but costly) approaches such as MD simulation [203]. Similarly, in this scenario a rapidly computable, heuristic scoring function would be preferable to a more accurate, but costly, physics-based FF. (3) Is the goal to predict the activity of a family of related small molecules (say Ļ, Ľ, Ł, Ḽ), at a particular binding site, in order to assess their value as potential lead compounds for drug development? This would require calculation of accurate binding free energies to protein $\mathcal{P}$ ($\Delta G^o_{bind}$ for $\mathcal{P}$•Ļ, $\mathcal{P}$•Ľ, $\mathcal{P}$•Ł, etc.). Similarly, a related goal might be to predict the effects of point mutants of $\mathcal{P}$, either engineered or naturally occurring, on the binding affinities for this set of compounds. This third scenario is the most computationally demanding, as accurate ligand-binding free energy calculations require extensive configurational sampling and an accurate FF representation of the physical interactions ([204] and references therein).

### 5.2.3. Basic principles, approaches

Docking consists of two parts: *(i)* a *sampling* method to general trial P•L structures (*poses*) and *(ii)* a *scoring* system to evaluate a pose by assigning it a value that presumably reflects its accuracy. Note that this is analogous to the basic approach in MD simulations, where the equations of motion serve as a sampling method (propagate the equations forward in time) and the FF serves the role of a scoring function. The task of sampling is also referred to as the 'search' problem in docking. The aim of accurate scoring is related to the goal of computing ligand-binding affinities. Modern docking research is dedicated largely to the sampling and scoring problems [110]. While many effective methodologies have been developed, many limitations continue to hamper the usage of docking in computer-assisted drug design (CADD) pipelines, in terms of *efficiency* (*coverage* – largely an issue of sampling) and *reliability* (*accuracy* – largely an issue of scoring functions). Further information on docking principles and approaches, including lists of software suites, have been reviewed in several places ([205-208, 77]). After outlining the demands on a docking method, the remainder of this subsection elaborates the two problems of sampling and scoring.

*The demands* — How we approach the sampling and scoring problems, e.g. what level of approximation is permissible, is dictated by the demands we make of a docking method for a specific application. For instance, the docking method used in a CADD pipeline will necessarily be cruder (computationally cheaper, per compound) than the techniques used in a careful study of binding energetics (using, for instance, free-energy perturbation calculations on a small set of ligand compounds). The demands of most docking applications occupy one of three levels: *(i)* At the crudest level, a docking study may simply aim to identify *active* ligands from a library of candidate compounds, even if the predicted P•L structure for that compound is incorrect or there are minor inaccuracies in the pose. Here, 'active' is taken to mean high-affinity binding (sub-μM), though in principle it simply means bio-active, irrespective of in vitro binding strength. In this context, experimental binding data from high-throughput screening can help cross-validate docking results, thereby improving the overall accuracy of the docking study. *(ii)* At a more demanding level, the docking approach will identify the 'true' ligand by discriminating it from a pool of inactive compounds and will also correctly predict the pose of this ligand in the binding site. At this level, crystallographic or NMR structures of the P•L complex (or a close analog P•L′) provide a means of validation that can also be used to refine the ligand design. *(iii)* At the highest level of stringency, a docking method will successfully identify true binders, correctly predict the P•L structure, and accurately estimate $\Delta G_{bind}^{o}$. At present, this level *(iii)* is not computationally feasible as part of a high-throughput pipeline because accurate free energy calculations require both extensive configurational sampling and an accurate scoring system, in the form of a physics-based FF that can account for binding-associated changes in entropy of the ligand and receptor, solvation effects, and so on [207].

*Sampling* — A docking *search method* is used to sample configurational space as efficiently as possible, thereby generating many P•L structures for scoring and ranking. To achieve this, three sets of issues must be considered: *(i)* the sampling algorithm, *(ii)* how molecular flexibility is treated, and *(iii)* whether the docking will be *blind* (unknown binding site) or *focused* on a particular region of the receptor (a known or suspected binding site). The challenge is clearly much greater in blind versus focused docking: As shown in Fig 6, a blind docking study must consider the entire solvent-accessible surface of the receptor in order to avoid false negatives, whereas in focused docking more extensive sampling, and therefore better docking, is possible because the same computational resources can be focused on a more limited spatial domain (finer grids, more exhaustive sampling of trial poses, cf. Fig 6A and B). In the absence of high-resolution structural information, lower-resolution experimental data, such as from chemical cross-linking, can greatly aid a docking study by enabling a focused calculation instead of blind docking. Determining the site for a focused docking study can be accomplished manually or by more automated methods, including MD simulations of the target protein with small probe molecules to identify binding sites [203].

Of the four possibilities for treating ligand and protein flexibility, $\{\mathsf{L},\mathsf{P}\}\times\{\text{flexible, rigid}\}$, virtually all current software suites treat the small-molecule L as flexible (rotatable single bonds), while some packages allow for partial inclusion of protein flexibility [77], e.g. by considering only a subset of residues centered near the presumptive binding site (the yellow receptor side-chains in Fig 6B). Given today's computing power and docking algorithms, what can be achieved lies between the two extremes: a rigid-L/rigid-P treatment is unnecessarily crude and inaccurate, but flexible-L/flexible-P is not yet routinely feasible (in the sense of a fully flexible protein, including all side-chains and backbone).

The type of sampling algorithm (issue *i*) and the treatment of flexibility (issue *ii*) are closely intertwined. Docking codes typically take one of three approaches to sampling: *(a)* systematic, *(b)* stochastic, or *(c)* simulation-based. Systematic methods pursue a brute-force calculation, wherein the geometric parameters of interest are systematically varied across all possible values of those DoFs. For example, if we are interested in torsion angles 1, 2, and 5 of L, then we might sample each of those torsion angles across the full angular range in increments of 5°. Even this unrealistically simple case yields $(360/5)^3 = 373{,}248$ combinations of parameter values to evaluate in the scoring function. Moreover, the previous figure severely underestimates the true number of potentially important degrees of freedom: 3 translational and 3 rotational DoF describe the rigid-body location and orientation of L with respect to P (these must be sampled with some reasonable granularity), and the number of torsional DoF in P dwarfs the above estimate for L. This combinatorial explosion in the dimensionality of the search space grows geometrically with the number of DoF and severely limits the general effectiveness of systematic sampling approaches. For these reasons, most docking codes utilize stochastic search methods such as Monte Carlo sampling, genetic algorithms, or 'tabu' search (these methods are described in [209, 206, 207]). An example of a simulation-based strategy would be to use MD-based simulated annealing to generate trial poses; an advantage of simulation approaches is that they offer a natural way to incorporate molecular flexibility in the docking calculation, but a disadvantage is the computational cost required for atomistic simulations to cross high-energy barriers and achieve reasonable sampling. Because of the difficulties of the search problem, many alternative sampling strategies have evolved. Most of these schemes incorporate a stochastic or simulation-based algorithm as a central routine. Examples for sampling the space of possible ligands/poses include incremental 'fragment-growth' methods [210] and database methods (libraries of pre-generated conformers that can be manipulated [211]). Examples of alternative approaches to treat receptor dynamics include the usage of protein side-chain rotamer libraries, protein ensemble grids, and so on (see reviews cited above).

*Scoring* — Reliable docking calculations require a robust scoring system, wherein numerical values are assigned to each of the candidate P•L complexes generated by the sampling algorithm. These numerical values (or 'scores') are often assumed to correspond to $\Delta G^{o}_{bind}$ values, even when such scores are more knowledge-based rather than physics-based. Early docking codes [212] assigned scores based on geometric fit/steric complementarity between P and L; such an approach successfully captures the essence of apolar interactions (and therefore works with hydrophobic ligands), but neglects potentially important effects such as electrostatic complementarity and the donor/acceptor directionality of H-bonds [201]. As described in the remainder of this section, modern scoring systems are either *(i)* FF–based, *(ii)* empirically-derived functions, or *(iii)* knowledge-based.

Force-field–based scores adopt the MM approach (§3.2, §4.3), with physically motivated terms to model the energetics of inter-atomic contacts between P···L. In fact, many FF-based scoring systems used in docking stem from the transferable FFs developed over the years for MD simulations (AMBER, CHARMM, etc.). While a disadvantage of the FF-based scoring systems is that they are computationally costly, versus the other two types of scoring approaches, an advantage is that their physical basis permits one to manipulate the terms in a comprehensible manner; for instance, one can 'soften' atomic interactions by adjusting the repulsive wall of the Lennard-Jones potential from $r^{-12}$ to $r^{-9}$.

The other two types of scoring systems, empirical and knowledge-based, are both statistical in nature. Empirical scoring systems utilize simpler functional forms than in the FF approach, with parameters that are obtained by fitting against experimentally determined binding affinities [213]. A strength of this approach is that its simpler functional forms are computationally cheaper to evaluate; a drawback is that the nonbonded interaction terms, because they are derived by statistical regression, could be physically rather *ad hoc* (they may not be transferable to other classes of ligands, the terms in the equation may be difficult to troubleshoot as they do not correspond to physicochemical properties, and so on). Also, as with many statistical fitting approaches, the parameters of the scoring system can be inadvertently over-trained against the necessarily limited datasets from which they are derived, thus limiting the transferability of the scoring approach [214, 215].

Knowledge-based scores derive an effective energy for an interatomic interaction A···B by computing the statistical occurrence of this interaction (e.g., frequency of A···B pairs) in a large database of known 3D structures. Implicit in this approach, which is based on the concept of a PMF (§4.7.2), is the assumption that all of the physics that might be relevant to an A···B interaction is implicitly contained as pairwise interactions in our databases of known 3D structures [216, 217]. As is the case with empirical scores, knowledge-based scores are rapidly evaluated because of their simple functional forms and limited number of terms. In addition to the issue of transferability and other caveats about statistical potentials, more subtle drawbacks to the knowledge-based approach include its basis in the concept of a reference state for the PMF (the reference state is a clear idea in the thermodynamics of simple systems, but a less clear concept when counting pairwise A···B interactions in a macromolecular complex). To make the calculation of scores numerically tractable, the scoring functions of the statistical potential are evaluated on a spatially discretized grid. That is, the ligand is positioned at successive points on a user-defined grid, possibly with sub-Å spacing between grid points (Fig 6). Docking codes achieve run-time efficiency by pre-computing these 'atomic affinity grids', which specify, for each unique atom type, the interaction between that atom and other atom types (such as may occur in the ligand). Such grids are computed for the various nonbonded components of the potential energy (e.g., Coulombic, vdW), and accelerate the overall calculation by obviating the need to re-compute the grid for each successive translation of the ligand across the grid [207].

The three scoring methodologies described above — FF-based, empirically-derived, and knowledge-based potentials — serve as a starting point for several strategies to enable more accurate scoring and ranking of docked poses. For instance, the core idea in the 'consensus scoring' approach is to combine for a single docking calculation the results obtained by application of different scoring schemes, parameters settings, etc., thus providing a consensus score for each pose. If the underlying inaccuracies of each scoring system are statistically independent of one another (a major assumption), then any such errors would cancel and the consensus score should serve as a more accurate predictor by which to rank poses in terms of binding affinities. In the 're-scoring' strategy [218, 219], the results from an initial docking calculation (i.e., the poses, rank-ordered by score) are refined by re-scoring the list of poses using a higher-accuracy (more costly) scoring scheme, such as the MMPBSA approach. The MMPBSA approach addresses the three chief shortcomings of most scoring systems – entropies, solvation, and electrostatics – by using a molecular mechanics-based approach to estimate conformational entropies (MM), a continuum treatment of electrostatics via the Poisson-Boltzmann equation (PB), and surface area terms (SA) to capture solvation effects [220].

### 5.2.4. Software packages

The key idea in docking is to rapidly generate many P•L trial structures and then evaluate each candidate using scoring functions such as those described above. Most of the variation between different docking packages stems from differences in how they address the sampling and scoring problems. The first general-purpose protein–ligand docking code (DOCK) was developed by Kuntz and coworkers at UCSF and released in the late 1970s [221]. In the intervening thirty years, a multitude of approaches have been developed and implemented as software suites that are either freely or commercially available. Because many heuristic approximations, empirical optimizations (parameter-tuning), and other computational 'shortcuts' enter these packages to make the calculations feasible, there can be great variation in the performance of different programs for different types of problems (e.g., blind versus focused docking), and with respect to different performance metrics. For example, a major performance criterion, in terms of sampling, is the treatment of flexibility. Virtually all modern software packages treat ligands as flexible, but until recently only few codes incorporated even partial receptor flexibility as a way to better sample the space of possible protein–ligand binding modes [77]. Because software packages rapidly evolve and algorithms are under continual development, the set of available docking codes, and their speed, accuracy and other performance metrics, are fast moving targets. Software suites are not listed here, as compilations of some of the most prevalent docking codes are available in the literature. For example, Kitchen *et al.* [206] and Sousa *et al.* [207] provide tables of docking programs, with the codes categorized by sampling approach, scoring methodology, handling of receptor flexibility, and various other criteria.

### 5.3. Protein–protein docking

Most cellular processes are mediated by protein-based assemblies [222, 5, 223], such as protein folding chaperones [224], polymeric components that form cytoskeletons [225], and ribonucleoproteins such as the ribosome [115]. For simplicity, consider only protein–protein interactions, and specifically the case of homotypic interactions of a protein '$\mathcal{P}$' that assembles into oligomers of *n* subunits, $\mathcal{P}_n$. The monomer, $\mathcal{P}$, may be non-functional, partly functional, or it may exhibit some unique, alternative function (apart from $\mathcal{P}_n$). In any case, the precise biochemical function of $\mathcal{P}$, such as binding a specific ligand signal, may be similar or dissimilar to the physiological function of the full oligomer in vivo; if $\mathcal{P}$ and $\mathcal{P}_n$ have similar biochemical properties, then the oligomer may simply act by presenting multiple interaction sites (a concept termed *avidity*). Most often for self-associating proteins, the biologically functional unit is the oligomeric assembly; it is this assembly which supplies some vital biochemical function and is therefore the evolutionarily conserved entity [223]. In such assemblies, *head → head* association of subunits yields complexes that are generally *closed*, whereas interactions with *head → tail* polarity can give either closed (cyclic) assemblies or open-ended 'runaway' structures (polymeric fibrils in one dimension, sheets or layers in two dimensions, and crystals in three-dimensions [226]). In all such cases, protein–protein docking can be applied.

Assume we know from experiments that well-defined homomeric A•A or heteromeric A•B associations occur in vitro. Such information is often accessible via analytical ultracentrifugation, FRET spectroscopy, or other solution-state biophysical approaches [227]. Then, protein docking can help further characterize these complexes by addressing basic questions: *(i)* Can a stable A•B complex be identified by the docking methods (plural, if trying a consensus docking approach)? *(ii)* If so, how many such distinct A•B binding modes are there? For example, are there two or three distinct binding patches, leading to various A•B geometries, or does a single geometry recur as the top hits in the docking trials? *(iii)* What is the predicted binding affinity for the A•B complex? How does this value compare to that determined from, e.g., isothermal titration calorimetry or surface plasmon resonance measurements? Though not always straightforward, such questions can be addressed via protein docking. Methods for protein docking have evolved in parallel with the protein–ligand field, albeit with a time-lag that is due, in part, to the relative scarcity of 3D structural data on protein–protein complexes versus protein–ligand complexes. Protein docking faces many of the same computational challenges as protein-ligand docking, with two specific types of problems taking on heightened significance in the protein-protein case: *(i)* protein flexibility should be treated, at least at the side-chain level, as numerous pairwise contacts between side chains define an A•B interface (the energetics of the binding process is at least partly governed by the loss of conformational entropy of these side-chains); *(ii)* the need to accurately model solvation becomes even more pronounced in protein docking, as desolvation of the interface is a major determinant of the association mechanism. As with protein-ligand docking, many computational strategies have been developed to address these questions; this active field has been reviewed recently [228, 215].

Protein–protein docking has taken on renewed relevance in this post-structural genomics era. We now have 3D structures of many of the isolated components of cellular complexes, but not the entire assemblies. Many such assemblies are only transiently stable, making them recalcitrant to structure determination via X-ray crystallography or NMR spectroscopy. Efficacious protein–protein docking, along with protein–nucleic acid and protein–ligand docking, would provide a path towards predicting the structures of such complexes and thereby bridge the rapidly widening gap between our knowledge of individual protein structures and the cellular-scale structures into which they assemble.

## 6. **Conclusions**

The interior of a cell is crowded with biopolymers, molecular assemblies, and small molecules. This dense environment is a highly dynamic network of molecular contacts (Fig 1A), meaning that a full understanding of any cellular pathway requires an accurate and detailed description of the molecular dynamics within and between its components. Though such interactions vary immensely in terms of possible types (chemical groups), strengths (thermodynamic stability), and lifetimes (kinetics), molecular simulations provide a powerful approach. The atomic contacts that mediate the binding of a small-molecule inhibitor to an enzyme active site are of the same physical nature as the contacts that stitch together the dozens of subunits in a cellular-scale assembly such as the ribosome. The fundamental interactions are the same, only the chemical variety and number of pairwise (and higher-order) contacts differ; these differences in molecular recognition give rise to the variation we see in biological assemblies. The difficulties in experimentally characterizing the conformational dynamics of biomolecular assemblies have driven advances in simulation-based approaches, such as MD and docking. The power of the simulation approach stems from its origin in statistical mechanics, which links the experimentally accessible macroscopic properties of a system to the microscopic structure and dynamics of its constituents. Indeed, the versatility of simulation-based approaches is immense, as molecular simulations have been applied to studies of *(i)* normal protein function (e.g., allostery), *(ii)* protein malfunction (aggregation diseases, mutations in metabolic diseases, etc.), *(iii)* protein structure prediction, design and engineering (e.g., homology modeling), *(iv)* macromolecular structure determination via crystallography, NMR and electron microscopy, and *(v)* structure-based drug design. The utility and applicability of molecular simulations will only continue to grow with our increasing knowledge of biological systems as highly dynamic arrays of molecular interactions.

## Acknowledgements

This work is dedicated to the memory of Aubin Mura. We thank RG Bryant, CT Lee, KK Lehmann, AD MacKerell, JA McCammon, and PS Randolph for discussions and feedback. Portions of this work were supported by the University of Virginia, the Jeffress Memorial Trust (J-971), and an NSF CAREER award (MCB-1350957).

## Side Boxes [Optimal location in the text is indicated.]

### Box 1: Notational conventions, abbreviations, symbols [located early, before §2]

- Words or phrases are italicized either for *emphasis* or when introduced as *new terminology*; vectors are indicated either in bold italics (e.g., $\boldsymbol{r}$ for the position vector) or by an arrow above the letter (e.g., $\vec{r}$).
- Abbreviations, acronyms, symbols: BD, Brownian dynamics; DoF, degree of freedom; FF, force-field; FFT, fast Fourier transform; LD, Langevin dynamics; MC, Monte Carlo; MD, molecular dynamics; MM, molecular mechanics; NMA, normal mode analysis; PBC, periodic boundary conditions; PCA, principal component analysis; p.d.f., probability distribution function; PME, particle-mesh Ewald; PMF, potential of mean force; QM, quantum mechanics; RMSD/F, root-mean-square deviation/fluctuation; a single center dot '•' indicates an intermolecular complex and a triple '···' denotes specific interatomic interactions
- The following symbols denote physical constants or frequently appearing quantities: $E_{tot}$, total system energy; $\mathcal{U}$, potential energy (also written $E_{pot}$ and known as the *internal energy* of a molecule); $\mathcal{K}$ or $E_{kin}$, kinetic energy; $T$, absolute temperature (Kelvin); $S$, entropy; $H$, enthalpy (or Hamiltonian, $\mathcal{H}$, depending on context); $A$, Helmholtz free energy; $G$, Gibbs free energy; $Z$, partition function; $m$, mass; $N_{\mathrm{A}}$, Avogadro constant (≈$6.02\times10^{23}$ entities/mole); $k_{\mathrm{B}}$, Boltzmann constant (≈$1.38\times10^{-23}$ J/K)

### Box 2: Simulation-related physical concepts and terminology [located in §3 (Physical principles)]

- *Ensemble*: A collection of $N$ particles possessing some well-defined, bulk thermodynamic properties, such as temperature ($T$), pressure ($P$), or mean energy ($E$); importantly, $T$, $P$, $E$, and all other macroscopic quantities become statistically well-defined, with only infinitesimal fluctuations about the mean, beyond $\sim10^5$ particles. Three ensembles commonly used in MD simulations are $NVE$ (*microcanonical*), $NVT$ (*canonical*), and $NPT$ (*isothermal–isobaric*), which correspond to fixed numbers of particles, volume, energy, etc., as indicated by the symbols for each. These three ensembles correspond to maximizing the system entropy, minimizing the Helmholtz free energy ($A = \mathcal{U} - TS$), or minimizing the Gibbs free energy ($G = H - TS$), respectively. For some types of systems (e.g., an ion channel in a planar membrane bilayer), less common ensembles may become useful (e.g., the constant surface tension [$\gamma$] and normal pressure [$P_\perp$] ensemble, $NP_\perp\gamma T$).
- *Phase space*: For a dynamical system of *N* particles, this is the multidimensional space of all values of position (***q***; *3N* DoF) and momenta (***p***; *3N* DoF). Importantly, proteins and other systems of interest are well-defined collections of particles (there is a particular pattern of covalent connectivity that defines, say, a leucine versus an isoleucine), so not all arbitrary values and combinations (***q***, ***p***) are allowed; also, particular regions of phase space are preferentially populated, and at equilibrium the Boltzmann distribution is the probability distribution function (p.d.f.) governing the population of these accessible regions of phase space. In short, phase space can be viewed as a hyper-dimensional inventory of all the potential microscopic states of a system together with the probability of occurrence of each; thus, as a concept phase space encompasses all that is knowable about the microscopic dynamics of a thermodynamic system.
- *Trajectory*: The list of coordinates ($\vec{r}_i$) and velocities ($\vec{v}_i$) for each atom $i$ in a system, as a function of time, over the course of a dynamics simulation. An individual structure from this time series $\{\vec{r}_i(t)\}$ is often referred to as a *snapshot* or *frame* from the trajectory.
- *Ergodicity*: This central axiom of statistical mechanics is that the *ensemble average* of some observable property (*A*) of a system, denoted $\langle A \rangle$, converges to the same value as the *time-average* of that property, denoted $\overline{A}$, in the limit of infinite sampling. This is the fundamental justification for applications of MD, as it stipulates that trajectory-averaged properties computed for a single molecule in isolation (a simulation system) equals the bulk thermodynamic properties of the system. This is also why sufficient sampling is crucial in MD, where 'sufficient' means to the point of convergence of bulk properties.

### Box 3: Overview of MD simulations [located in §4 (MD simulations)]

- *What is it?* — A computational method to numerically evaluate the equations of motion for a set of particles, such as the atoms in a protein. The result is an MD trajectory, which is a detailed description of the dynamics of the system on the timescale of the simulation.
- *How is it done?* — The equations of motion for such a complex system are not soluble, neither in principle (many-body problem) nor in practice (analytically intractable to solve for dynamics of $6N$ degrees of freedom, where $N$ may exceed $10^3$ non-hydrogen atoms in a small-sized protein). Instead, we discretize time and numerically integrate the equations of motion via a finite difference method: Given a set of initial positions ($\boldsymbol{r}_i(t_n)$) and velocities ($\boldsymbol{v}_i(t_n)$) for each particle $i$ at step $n$ (time $t_n$), compute the forces on each atom (from the gradients with respect to the force-field potential) to obtain accelerations. Next use the positions ($\boldsymbol{r}_i$), velocities ($\boldsymbol{v}_i$), and accelerations ($\boldsymbol{a}_i$) with the classical equations of motion to obtain updated positions and velocities for step $n + 1$ (= time $t_n + \delta t$, where $\delta t$ is the integration timestep, typically ~1-2 fs for biomolecular simulations).

**Box 4: Concepts and terminology: Force-fields** [located in §4 (MD simulations)]

The following terminology often appears in connection with force-fields:

- *Additivity*: If the forces and energetics of the interaction between two particles, *A* and *B*, are not influenced by the presence of a third particle, *C*, then the interaction is said to be *additive*; in this case, because we are considering pairs of particles, the forces are described as *pairwise additive*.
- *Polarizability*: The susceptibility of the electronic distribution about an atomic nucleus to distortion by an external electrical field, such as may arise from neighboring charged groups. This can be an important effect in highly-charged systems such as nucleic acids. Until recently, polarizability has been almost always neglected in MD force-fields and simulations, as its inclusion makes the MD calculation more costly.
- *Transferability*: In FF development, this is the idea that the physicochemical parameters developed for so-called *model compounds* (e.g., a blocked alanine) can be *transferred*, without loss of validity or accuracy, to chemically related compounds (e.g., an alanine residue in a polypeptide); such parameters are typically derived via high-level QM calculations that are feasible only for small model compounds. The notion of transferability is fundamental to the development of generalized force-fields.
- *Water model*: The precise geometric structure (bond lengths, angles) and electronic structure (e.g., location and magnitude of partial charges) used to represent a $H_2O$ molecule, as well as the types of physical effects included in the treatment (e.g., polarizability). Several water models have been developed over the years (TIP3P, SPC, etc.); the main differences between them concern the number of 'interaction sites' (e.g. lone-pairs as dummy sites), how structural flexibility/rigidity is handled, and how water molecule polarizability is treated.

**Box 5: Concepts and terminology: Docking** [located in §5 (docking)]

The following terminology often appears in the docking literature:

- *CADD*, *SBDD*: These acronyms are common in the docking literature, and denote *computer-aided drug design* and *structure-based drug design*; protein–ligand docking is a key step in most CADD workflows. A related concept is *HTS* (*high-throughput screening*), which may be performed experimentally (via robotic automation) or computationally (*virtual screening* of candidate drug compounds or other small ligands via in silico protein–ligand docking pipelines).
- *Receptor/ligand*: In a binary interaction, P•L, the larger entity (typically a protein or nucleic acid) is known as the *receptor* and the smaller molecule, such as a drug compound, is known as the *ligand*; analogous terms from chemistry are *host* (receptor) and *guest* (ligand). In drug-design applications, ligands that bind a receptor and elicit a positive response are known as *agonists*, whereas *antagonists* bind and inactivate receptors.
- *Pose*: The geometry or *binding mode* of a ligand in a receptor binding/active site. The pose is precisely described via *(i)* the usual six DoF that specify the rigid-body location of the ligand in space (three translational + three orientational parameters, relative to the receptor), and *(ii)* the exact 3D structure (*conformation*) of the receptor-bound ligand, in terms of its internal DoF. Typically only torsion angles (1-4 interactions) for the ligand, and possibly for receptor residues lining the active site, need to be considered, as bond lengths and angles do not significantly deviate from their standard reference values at physiological temperatures.
- *Pharmacophore*: A 3D model that defines, for a specific class of receptors, the important features of cognate ligands. Distinct chemical regions of the ligand are described in terms of physicochemical properties, including the relative contributions of each region and its associated properties to ligand-binding energetics and geometry. The development of *dynamical pharmacophores* is a modern research direction that aims to transcend static models by accounting for ligand flexibility, thereby improving pharmacophore-based methods for drug discovery and modeling of dynamic molecular interactions.

## Figure Captions

**Figure 1: Molecular interactions over many length-scales**. Structural biology and molecular simulations have reached the point that atomically-detailed models can now be built for the bacterial cytoplasm, and dynamics in this crowded medium can be studied. A snapshot from such a simulation is shown in (A); as implied by this image (from [3]), a cell can be defined by its set of molecular interactions. Flexibility in the number (few, dozens, hundreds) and types (polar, hydrophobic, etc.) of contacts yields immense variability in the resulting complexes. For instance, panel (B) shows part of the structure of the bacterial ribosome (protein blue, RNA yellow) bound to the antibiotic chloramphenicol (vdW spheres near center). This cellular-scale assembly is a vast network of protein⋯RNA (pink lines), protein⋯small-molecule (green lines) and protein⋯protein (not shown) contacts. Of these, protein–ligand interactions are the simpler to treat (local length-scale, fewer contacts) and are also of major pharmaceutical relevance, as enzyme inhibitors and other drugs are often small organic compounds. As an example of such interactions, panel (C) shows the anticancer drug imatinib bound to the tyrosine kinase ABL2 (an oncoprotein associated with several cancers). As is true of many ligand-binding sites, the compound binds in a concave, cleft-like region on the solvent-accessible protein surface [229]. The exact location of this binding site — between two protein domains (SH2 domain in gray, kinase domain in brown) — is related to imatinib's inhibitory potency (imatinib⋯ABL2 interactions block ABL2's phosphorylation activity). Myriad molecular interactions similar to the ones shown here are forming, persisting, or dissociating in a cell at any given moment.

**Figure 2: Biomolecular interactions and dynamics: Relevant timescales**. Biomolecular structure is modulated by dynamical processes that span several decades, ranging from ps–scale side-chain rotations to much longer (≈μs) times for rigid-body translation and rotation of higher-order structural units. Secondary structural elements, super-secondary structural elements (e.g., helix-turn-helix motif or a $\beta$-hairpin), or entire protein domains can engage in 'collective motions' on even longer timescales. Though omitted from this schematic for clarity, distinct motional modes also occur in nucleic acids, such as ns–scale re-puckering of nucleoside sugar rings and the longer characteristic times for global twisting, stretching, and bending of duplex helices. The terminology often used to describe these dynamical regimes includes *ultrafast* (≤ fs), *fast* (~fs ↔ ~ps), *infrequent* (~ps ↔ ~ns), and *intrinsically complicated* (~μs ↔ ~ms) processes. As a point of reference, the $\delta t \sim 1$-fs integration step used in most atomistic MD simulations is indicated. The approximate year in which simulations of a given duration (ps, ns, …) became at least feasible, if not routine, is shown above the timeline; for instance, μs-scale simulations became computationally attainable (multiple such simulations began appearing) shortly after 2005.

**Figure 3: Phase space and its sampling via MD and MC**. (A) A diagram of phase space for the simple harmonic oscillator, taken as a one-dimensional spring with a mass $m$ attached. This dynamical system is described by the potential $\mathcal{U}(x) = -½k(x - x_0)^2$, where $x$ is the coordinate of the mass, $x_0$ is its relaxed (equilibrium) position, and $k$ the spring constant ($k \sim$ stiffness). Differentiation of this equation yields the force $F(x) = -k(x - x_0)$, which we can analytically solve for the values of position and momenta as shown in (A); the position is labeled by a '$q$', rather than '$x$,' in panel (A) because '$q$' is often used to indicate a generalized coordinate in classical mechanics, and is the same as $x$ for the simple case of a one-dimensional harmonic oscillator. Consistent with our intuitive notion of oscillatory motion of a spring, note that *(i)* the mass reaches a minimal velocity (=0) at the two 'turning points' of maximal and minimal compression of the spring ($(q, p) = [\pm q_{max}, 0]$), and *(ii)* this dynamical system traces a repetitive *orbit* in phase space. The phase space of a more complex dynamical system (e.g. a protein with *N* atoms) is inordinately more complicated – it consists of $(q, p)^{3N-6}$ dimensions, and trajectories in this space may be irregular (not periodic). Exploring such a hyper-dimensional phase space requires some form of conformational sampling. Two well-established sampling approaches are MD simulations and Monte Carlo (MC). What is the difference between these methods? MD aims to *simulate*, with physical realism, the actual motion of the particles in a system; as described in §4, this is done by integrating the equations of motion to propagate the atomic coordinates along a trajectory in the system's phase space ($t_1 \rightarrow t_2 \rightarrow t_3 \cdots$ in (B)). In contrast, MC proceeds as a series of discrete 'trial moves' (e.g., "*flip torsion angle 42 by 180°*"). The sequence of trial moves are independent of one another, and are accepted or rejected by comparison of the Boltzmann-weighted probability to a randomly generated number. Whereas MD is analogous to a game of connect-the-dots in phase space, MC can be thought of as skipping dot-to-dot in this hyper-dimensional space.

**Figure 4: Molecular interactions and force-fields, in context**. The core elements of a molecular mechanics-based FF, such as is used in MD simulations, are shown in the context of an important molecular interaction: the binding of the cancer therapeutic imatinib to ABL2 kinase (see Fig 1C). The overall structure of the ABL2•imatinib complex shows the location of the drug (ball-and-stick and semi-transparent vdW spheres); protein side-chains that interact at the binding site (ball-and-stick) are shown in atomic-level detail in panel (A). The atomic interac-

tions between ABL2⋯imatinib (B) include polar contacts (yellow dashes) such as hydrogen bonds, interactions that are more strongly electrostatic in character ($\delta^{+} \cdots \delta^{-}$), and numerous vdW interactions between non-polar groups of atoms (not shown for clarity). The components of a typical FF are schematically drawn in (B), showing the roles of these inter-atomic interactions (bond angle bending, torsional rotations, etc.) in mediating the molecular recognition process. In classical MD, the full potential energy ($\mathcal{U}$) is taken as a sum of various types of physicochemical interactions (shown in (B)), and each type of interaction is treated explicitly via specific terms in the FF equation (see text, Eq 5). The terms in Eq 5 account for *(i)* bond stretching (1-2 interactions), angle bending (1-3 interactions) and torsion angle rotation, as well as *(ii)* non-bonded interactions between apolar groups (a Lennard-Jones potential to model dispersive interactions). The short-range component of electrostatic interactions between fixed partial charges is modeled via Coulomb's law, and long-range electrostatics across the PBC lattice are treated via Ewald summation.

**Figure 5: MD simulations in a nutshell**. MD simulation is a multi-stage process that employs several chemical, physical, and computational principles (A). Working *left → right* in panel (A), a starting 3D structure is prepared by addition of solvent and other moieties, giving an initial list of atomic coordinates (time $t_0$). These 3D positions, together with the covalent chemical structure of the molecule, define the molecular system (blue box). Literally all of the precise atomic details that define a complete, solvated biopolymer structure are contained in the chemical *topology file*: the standard amino acids, nucleotides, common ions, the detailed patterns of covalent connectivity and the orders of bonds — which atoms are bonded to one another, various hybridizations ($sp^2$, $sp^3$), whether an amine is 1°, 2°, and so on. Force-fields also include a *parameter file* associated with the topology file, defining the functional form of the potential energy equation (Fig 4B), as well as the reference values for bond lengths ($r_0$), angles ($\theta_0$), multiplicity and phase of torsional angles, Lennard-Jones parameters, and so on. Conceptually, the three parts of an MD simulation system (green box) are *(i)* the force-field (gray box; not system-specific); *(ii)* the atomic positions and velocities over time (which are system-specific, and which give the MD trajectory); and *(iii)* the key details that describe the simulation to be performed – which thermodynamic ensemble, whether periodic boundary conditions (B) are employed, cutoff lengths for evaluation of non-bonded interactions, and so on. Taking this system as input, the MD software (black box) computes the forces on each atom from the gradient of the potential, $-\nabla\mathcal{U}(\vec{r})$. Newton's second law relates these forces to the acceleration of each atom, $\partial^2\vec{r}/\partial t^2$, which, in turn, relates to the atomic position and velocity by classical mechanics (C). The MD engine generates a trajectory by discretizing time, often with an integration step $\delta t \approx 1$ fs, and integrating the equations of motion. To achieve this, the algorithm iterates over all inter-atomic interactions (bonded and nearby nonbonded pairs), computes the forces of atoms on one another, and then uses these forces to update the positions and velocities of each atom, via numerical methods such as the 'leapfrog' integrator (C; see §4.4).

**Figure 6: Protein–ligand docking in action: A computed grid**. Many protein-ligand docking algorithms employ a discrete spatial grid over which the calculation is performed, as explicitly shown here. In this example, the receptor is the ABL2 tyrosine kinase and the ligand is the inhibitory compound imatinib (see also Fig 1C). Coarse grids were used for 'blind' docking over the entire receptor (A), while finer grids could be applied for more 'focused' docking centered on the (known) ligand-binding site (B). Docking grids were computed using AUTODOCK, and the illustration was created and rendered in PyMOL.

**Figure 1**: Molecular interactions over many length-scales

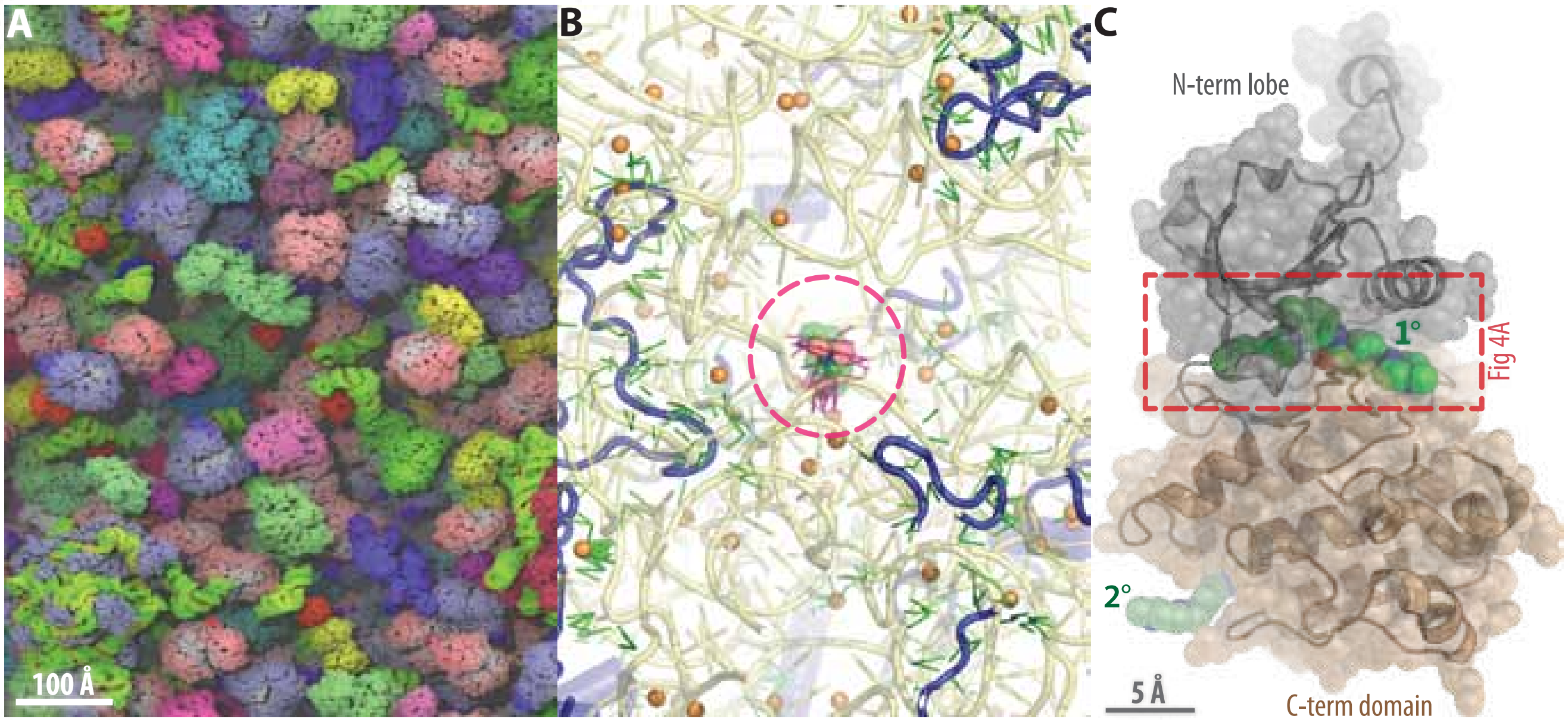

**Figure 2**: Biomolecular interactions and dynamics: Relevant timescales

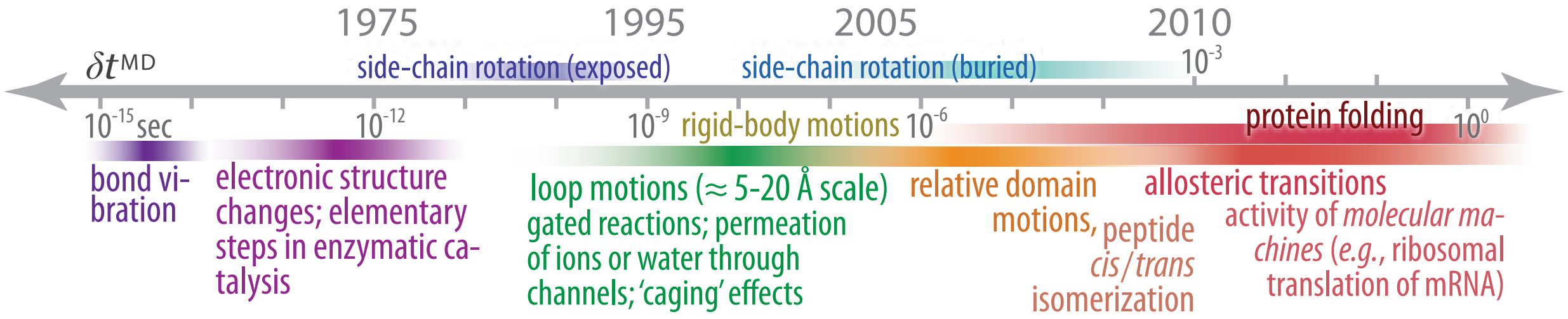

**Figure 3**: Phase space and its sampling *via* MD and MC

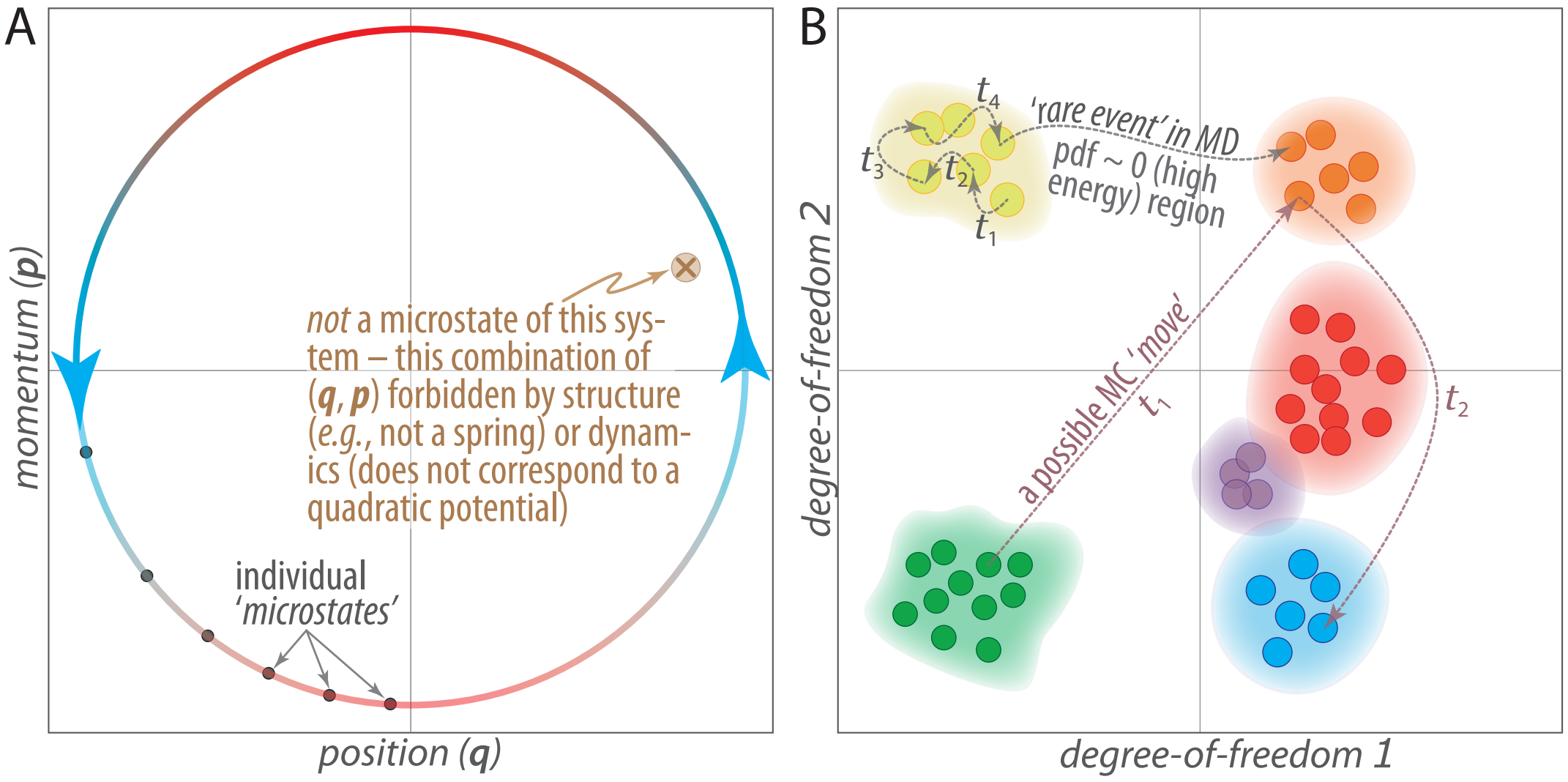

**Figure 4**: Molecular interactions and force-fields, in context

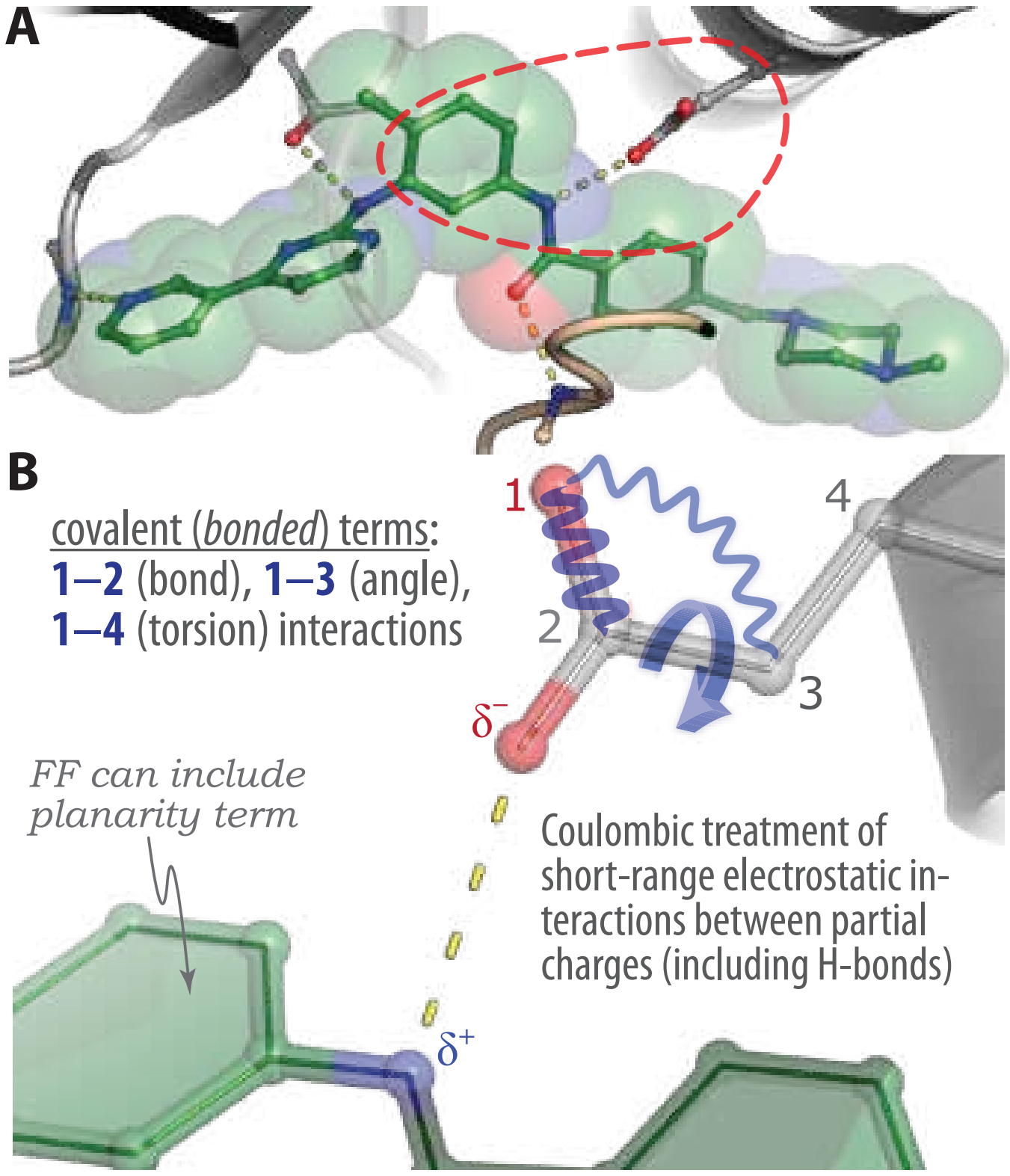

**Figure 5**: MD simulations in a nutshell

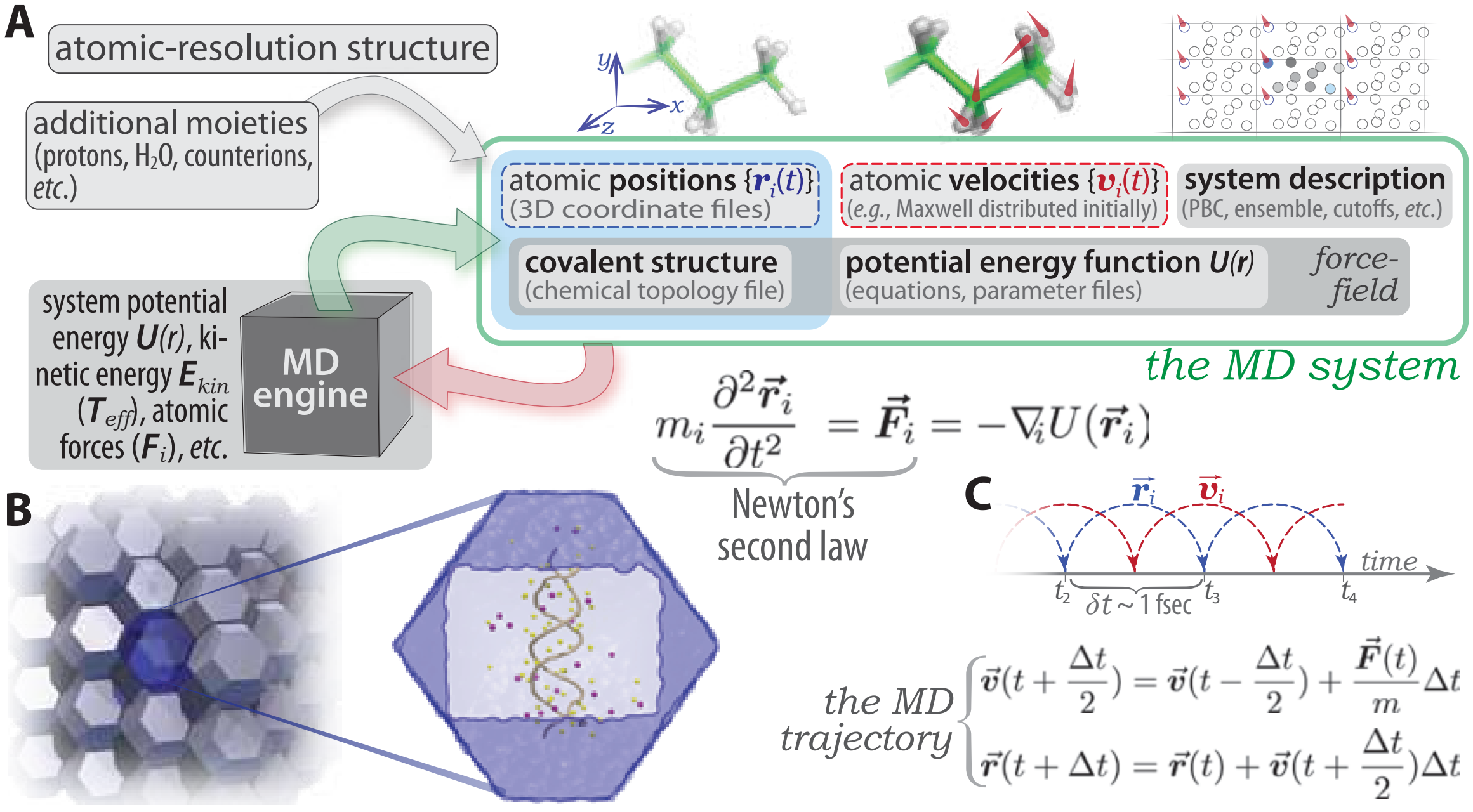

$$m_i \frac{\partial^2 \vec{r}_i}{\partial t^2} = \vec{F}_i = -\nabla_i U(\vec{r}_i)$$

$$\vec{v}(t+\frac{\Delta t}{2}) = \vec{v}(t-\frac{\Delta t}{2}) + \frac{\vec{F}(t)}{m}\Delta t$$

$$\vec{r}(t+\Delta t) = \vec{r}(t) + \vec{v}(t+\frac{\Delta t}{2})\Delta t$$

**Figure 6**: Protein–ligand docking in action: A computed grid

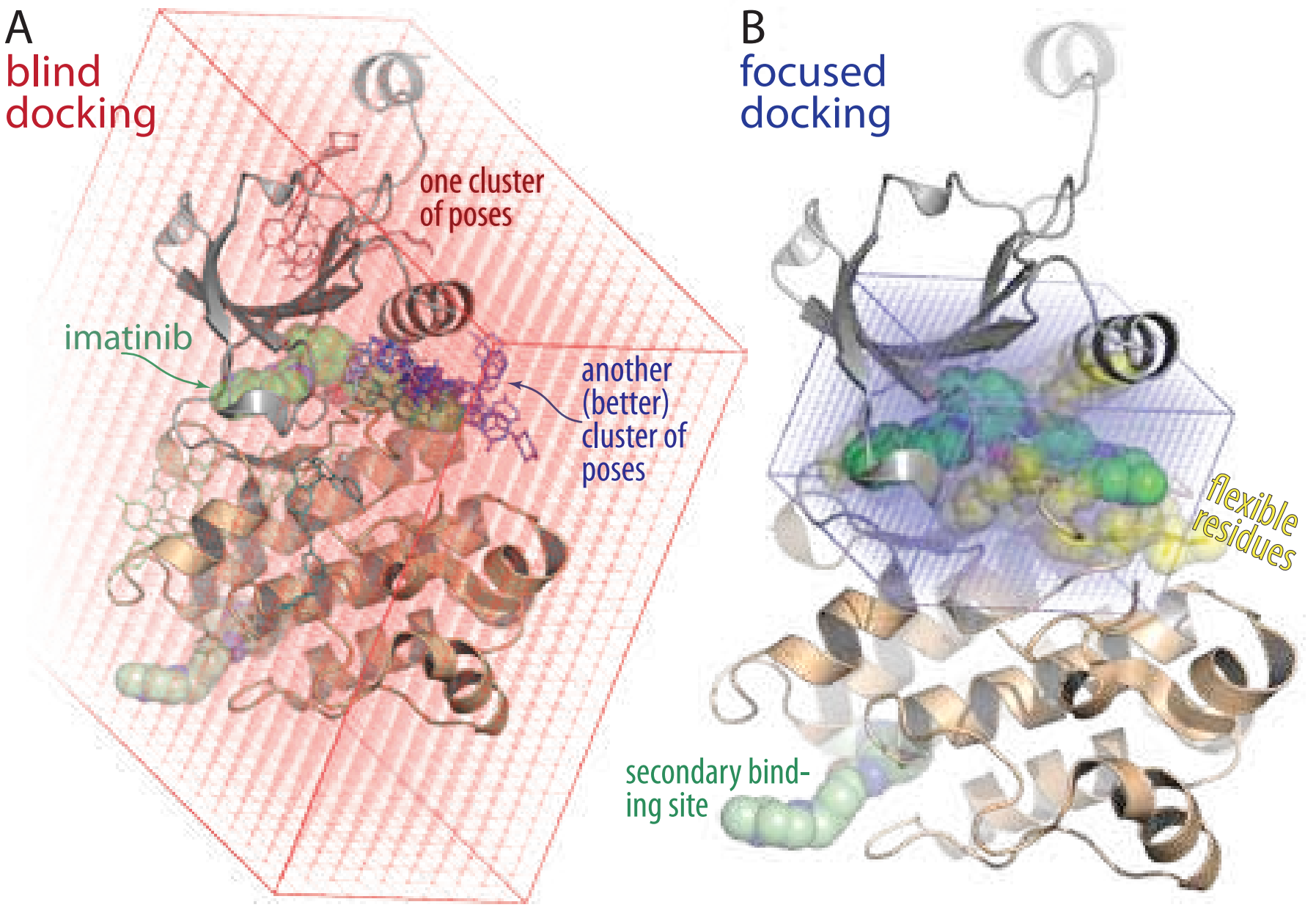